\documentclass{emulateapj}

\usepackage{psfig}
\usepackage{graphicx}
\usepackage{natbib}

\newcommand{\sgr}{\mbox{SGR\,J$1550-5418$~}}
\newcommand{\sgrnos}{\mbox{SGR\,J$1550-5418$}}
\newcommand{\degrees}{\ensuremath{^\circ}}

\shorttitle{\sgr bursts detected with the  {\it Fermi}/GBM}
\shortauthors{van der Horst et al.}

\begin{document}

\title{\sgr bursts detected with the {\it Fermi} Gamma-ray Burst Monitor during its most prolific activity}

\author{A.~J.~van~der~Horst\altaffilmark{1,2},
C.~Kouveliotou\altaffilmark{3},
N.~M.~Gorgone\altaffilmark{4},
Y.~Kaneko\altaffilmark{5},
M.~G.~Baring\altaffilmark{6},
S.~Guiriec\altaffilmark{7,8},
E.~G\"o\u{g}\"u\c{s}\altaffilmark{5},
J.~Granot\altaffilmark{9,10,11},
A.~L.~Watts\altaffilmark{2},
L.~Lin\altaffilmark{5,12,7},
P.~N.~Bhat\altaffilmark{7},
E.~Bissaldi\altaffilmark{13},
V.~L.~Chaplin\altaffilmark{7},
M.~H.~Finger\altaffilmark{1},
N.~Gehrels\altaffilmark{8},
M.~H.~Gibby\altaffilmark{14},
M.~M.~Giles\altaffilmark{14},
A.~Goldstein\altaffilmark{7},
D.~Gruber\altaffilmark{13},
A.~K.~Harding\altaffilmark{8},
L.~Kaper\altaffilmark{2},
A.~von~Kienlin\altaffilmark{13},
M.~van~der~Klis\altaffilmark{2},
S.~McBreen\altaffilmark{15},
J.~Mcenery\altaffilmark{8},
C.~A.~Meegan\altaffilmark{1},
W.~S.~Paciesas\altaffilmark{7},
A.~Pe'er\altaffilmark{16},
R.~D.~Preece\altaffilmark{7},
E.~Ramirez-Ruiz\altaffilmark{17},
A.~Rau\altaffilmark{13},
S.~Wachter\altaffilmark{18},
C.~Wilson-Hodge\altaffilmark{3},
P.~M.~Woods\altaffilmark{19},
R.~A.~M.~J.~Wijers\altaffilmark{2}
}

\altaffiltext{1}{Universities Space Research Association, NSSTC, Huntsville, AL 35805, USA}
\altaffiltext{2}{Astronomical Institute "Anton Pannekoek," University of Amsterdam, Postbus 94249, 1090 GE Amsterdam, The Netherlands}
\altaffiltext{3}{Space Science Office, VP62, NASA/Marshall Space Flight Center, Huntsville, AL 35812, USA}
\altaffiltext{4}{Connecticut College, New London, CT 06320, USA}
\altaffiltext{5}{Sabanc\i~University, Orhanl\i-Tuzla, \.Istanbul 34956, Turkey}
\altaffiltext{6}{Department of Physics and Astronomy, Rice University, MS-108, P.O. Box 1892, Houston, TX 77251, USA}
\altaffiltext{7}{University of Alabama, Huntsville, CSPAR, Huntsville, AL 35805, USA}
\altaffiltext{8}{NASA Goddard Space Flight Center, Greenbelt, MD 20771, USA}
\altaffiltext{9}{Racah Institute of Physics, The Hebrew University, Jerusalem 91904, Israel}
\altaffiltext{10}{Raymond and Beverly Sackler School of Physics \& Astronomy, Tel Aviv University, Tel Aviv 69978, Israel}
\altaffiltext{11}{Centre for Astrophysics Research, University of Hertfordshire, College Lane, Hatfield, AL10 9AB, UK}
\altaffiltext{12}{National Astronomical Observatories, Chinese Academy of Sciences, Beijing, 100012, China}
\altaffiltext{13}{Max Planck Institute for extraterrestrial Physics, Giessenbachstrasse, Postfach 1312, 85748, Garching, Germany}
\altaffiltext{14}{Jacobs Technology, Inc., Huntsville, AL, USA}
\altaffiltext{15}{University College, Dublin, Belfield, Stillorgan Road, Dublin 4, Ireland}
\altaffiltext{16}{Harvard-Smithsonian Center for Astrophysics, Cambridge, MA 02138, USA}
\altaffiltext{17}{Department of Astronomy and Astrophysics, University of California, Santa Cruz, CA 95064, USA}
\altaffiltext{18}{IPAC, California Institute of Technology, 1200 E. California Blvd., MS 220-6, Pasadena, CA 91125, USA}
\altaffiltext{19}{Corvid Technologies, 689 Discovery Drive, Huntsville, AL 35806, USA}
\email{Alexander.J.VanDerHorst@nasa.gov}

\begin{abstract}
We have performed detailed temporal and time-integrated spectral analysis of 286 bursts from \sgr 
detected with the {\it Fermi} Gamma-ray Burst Monitor (GBM) in January 2009, 
resulting in the largest uniform sample of temporal and spectral properties of \sgr bursts. 
We have used the combination of broadband and high time-resolution data provided with GBM 
to perform statistical studies for the source properties. 
We determine the durations, emission times, duty cycles and rise times 
for all bursts, and find that they are typical of SGR bursts. 
We explore various models in our spectral analysis, 
and conclude that the spectra of \sgr bursts in the $8-200$~keV band are equally well described by 
optically thin thermal bremsstrahlung (OTTB), a power law with an exponential cutoff (Comptonized model), 
and two black-body functions (BB+BB). 
In the spectral fits with the Comptonized model we find a mean power-law index of $-0.92$, 
close to the OTTB index of $-1$. 
We show that there is an anti-correlation between the Comptonized $E_{\rm{peak}}$ and the burst fluence and average flux.  
For the BB+BB fits we find that the fluences and emission areas of the two blackbody functions are correlated. 
The low-temperature BB has an emission area comparable to the neutron star surface area, 
independent of the temperature, 
while the high-temperature blackbody has a much smaller area 
and shows an anti-correlation between emission area and temperature. 
We compare the properties of these bursts with bursts observed from other SGR sources 
during extreme activations, and discuss the implications of our results in the context of magnetar burst models.
\end{abstract}


\section{Introduction}

Magnetars are members of the diverse neutron star family: 
they are isolated neutron stars whose persistent X-ray emission and soft gamma-ray bursts 
are powered by their extremely strong magnetic fields of $10^{14}-10^{15}$~G 
\citep{duncanthompson1992,kouveliotou1998,thompson2002}. 
Historically two classes of sources have been identified with characteristics that could be explained 
by the magnetar model: Anomalous X-ray Pulsars (AXPs) and Soft Gamma Repeaters (SGRs).
These original classifications were based on several distinct differences between the two populations \citep[for reviews, see][]{woodsthompson2006,mereghetti2008}, but it has now become clear that they share several properties. 
Therefore, AXPs and SGRs form one source class 
with small differences still maintained, 
most notably the rates and energetics of bursts during active episodes, 
with the luminosity distributions of AXP bursts being lower by orders of magnitude than those of SGR bursts. 

The source discussed in this paper was discovered by the {\it Einstein} X-ray satellite 
and named 1E\,1547.0-5408 \citep{lambmarkert1981}; 
it was subsequently confirmed in the {\it ASCA} Galactic plane survey \citep{sugizaki2001}. 
{\it Chandra} and {\it XMM-Newton} observations identified the source as a magnetar candidate 
based on its varying X-ray flux, spectral characteristics and the tentative association 
with the supernova remnant G\,327.24-0.13 \citep{gelfandgaensler2007}, 
although no X-ray pulsations or soft gamma-ray bursts were detected. 
The magnetar nature of the source was confirmed by the discovery of radio pulsations 
with the Parkes radio telescope, which were subsequently associated with 1E\,1547.0-5408 with ATCA observations; 
the radio source was named PSR\,J1550-5418 \citep{camilo2007}. 
It is one of only three magnetars exhibiting pulsed radio emission, the other ones being XTE\,J1810-197 \citep{camilo2006} and PSR\,J1622-4950 \citep{levin2010}. 
A surface dipole magnetic field strength of $2.2\times10^{14}$~G has been inferred from its spin period of $2.07$~s (the shortest of all magnetars) and its period derivative of $2.32\times10^{-11}$~s~s$^{-1}$. 
Further studies of the radio and persistent X-ray emission showed that the rotation 
and magnetic axes of this neutron star are nearly aligned \citep{halpern2008,camilo2008}. 

Given its characteristics and the lack of bursting behavior, the magnetar source was originally classified as an AXP \citep{camilo2007}. 
However, in 2008 the source started a series of three active episodes that
covered over half a year: the first was in October 2008, the second, most burst active one started in January 2009, and the last in March 2009. 
During the first episode, the {\it Swift} Burst Alert Telescope \citep[BAT;][]{krimm2008a,krimm2008b,israel2010} 
and the {\it Fermi} Gamma-ray Burst Monitor \citep[GBM;][]{vonkienlinbriggs2008,vanderhorstbriggs2008} 
detected several tens of bursts; as a result the source was reclassified as an SGR and designated \sgrnos. 
This reclassification was firmly established by the extreme bursting activity displayed on January 22, 2009, 
when hundreds of bursts were detected in one day by several instruments, 
i.e., {\it Swift} \citep{gronwall2009,scholz2011}, {\it Fermi}/GBM \citep{connaughtonbriggs2009,vonkienlinconnaughton2009}, 
{\it INTEGRAL}/SPI-ACS \citep{savchenko2009,mereghetti2009a,mereghetti2009b} and IBIS/ISGRI \citep{savchenko2010}, 
{\it Suzaku} WAM \citep{terada2009}, RHESSI \citep{bellm2009}, 
and KONUS-WIND \citep{golenetskii2009a,golenetskii2009b,golenetskii2009c}. 
These bursts had durations and luminosities typical of SGR bursts; in fact 
some were so bright that they caused ionospheric disturbances 
observed in very low frequency radio wave data \citep{chakrabarti2009,tanaka2010}. 

The distance to the source is currently uncertain, varying from $\sim9$~kpc, based on the radio dispersion measure \citep{camilo2007}, 
down to $\sim4-5$~kpc, based on the tentative supernova remnant association \citep{gelfandgaensler2007} 
and the detection of expanding dust-scattering X-ray rings \citep{tiengo2010}. 
The former estimate has large uncertainties arising from the Galactic free electron density model of \citet{cordeslazio2002} 
used for calculating the radio dispersion measure. Therefore, we adopt here a source distance of 5~kpc. Throughout this paper we use the notation $d_5=\rm{distance}/(5\,\rm{kpc})$, when calculating burst energies and luminosities. 

In this paper we present the temporal and time-integrated spectral analysis of 286 \sgr bursts detected with GBM in January 2009 
for which we have high time-resolution data. 
Given the very large field of view (the entire un-occulted sky) of GBM, 
this is the largest uniform sample of temporal and spectral properties of \sgr bursts during the January activation. 
This data set enables us to determine their durations, spectral shapes and energetics, 
and compare them with other magnetar sources during similar high bursting activity periods: 
SGR\,1900+14 \citep{gogus1999,israel2008}, SGR\,1806-20 \citep{gogus2000}, AXP\,1E\,2259+586 \citep{gavriil2004}, and SGR\,1627-41 \citep{esposito2008}. We also compare our results with the analysis of bursts detected by {\it Swift} during the same time period \citep{scholz2011}, and discuss them in the framework of current theoretical models. 
Here we focus on the January 2009 bursts, while the analysis of the October 2008 and February - March 2009 bursts are discussed in von Kienlin et~al. (in preparation) and Younes et al. (in preparation). The discovery of enhanced persistent emission in the GBM data during the beginning of activity on January 22 has been presented in \citet{kaneko2010}. 

In Section~\ref{sec:sample} we present our data selection of GBM bursts, and in Sections~\ref{sec:temp} and~\ref{sec:spec} we present our temporal and spectral analysis, respectively. We discuss our results and their interpretation in the context of other bursting magnetar sources 
in Section~\ref{sec:disc}, and we summarize in Section~\ref{sec:conc}.

\section{Sample and data selection}\label{sec:sample}

GBM comprises 14 detectors with a field of view of 8~sr (the entire sky un-occulted by the Earth): twelve NaI detectors covering a spectral range from 8~keV to 1~MeV, and two BGO detectors covering $0.2-40$~MeV
\citep[see][for an overview of the instrument and its capabilities]{meegan2009}. 
We use here only data from the NaI detectors because SGR bursts are not detected above 200~keV. 

GBM collects three data types: Time-Tagged Event (TTE), CTIME and CSPEC \citep{meegan2009}.
To perform detailed temporal and spectral analyses over the typical burst durations of $\sim0.1$~s, we selected for our sample only events with TTE data, which provide a time-tagged photon event list with a temporal resolution as low as 2~$\mu$s in 128 energy channels logarithmically spaced over the energy range of the NaI detectors.
The other two data types cannot be used for our purposes, 
because either their accumulation times are much longer (CSPEC) or they have a worse spectral resolution (CTIME).  
The TTE data type covers the time range from 30 seconds prior to 300 seconds after a trigger, 
and GBM has been designed to not trigger for 600 seconds after a previous trigger. This leaves at least 270\,s between consecutive triggers without TTE data. Fortuitously, \sgr triggered GBM very frequently (41 times on January 22 alone), providing
a good TTE data coverage for hundreds of bursts.

To perform a comprehensive and systematic study of the \sgr bursts, we applied our untriggered burst search algorithm 
to the CTIME data type in both continuous (0.256\,s time resolution) and trigger (0.064\,s time resolution) modes \citep{kaneko2010}. 
For each burst we required a count rate above background in the $10-50$~keV energy range 
of at least $5.5\sigma$ and $4.5\sigma$ in the first and second brightest detectors, respectively. Note that the GBM trigger criteria differ from the ones applied in our untriggered burst search, resulting in omitting several GBM triggers from our sample. 
We then compared each burst candidate light curve and spectral behavior to those of typical SGR bursts, 
and checked if the \sgr location was consistent with the relative rates in the NaI detectors at the time of the burst 
(for the orientation of the spacecraft at that time). 
The untriggered burst search resulted in 555 events on January 22 alone and 597 events in total up to January 29. 
This total number, however, includes events for which TTE data are not available and thus excluded from this analysis. 
Moreover, in several occasions multiple events were part of the same burst as defined by the method employed by \citet{gogus2001}. 
More specifically, we required that for two events to qualify as two separate bursts, the time between their peaks in the TTE data had to be greater 
than a quarter of the spin period of the SGR (0.5~s in the case of \sgrnos), 
and the count rate level had to drop to the background level between the peaks. 
Applying these criteria to our initial list of events, we collected 286 bursts from January 22 to 29, which is the sample  
we used for our detailed temporal and spectral analysis. 
  
There is a caveat, however, associated with our sample completion. For a significant fraction of the time we were not able to detect \sgr with GBM: during {\it Fermi} passages through the South Atlantic Anomaly (SAA) and when the source was occulted by the Earth. 
Specifically, on January 22 {\it Fermi} was in the SAA for 3.5 hours and \sgr was occulted by the Earth for almost 8 hours. 
Since some of these times overlap, the total time that GBM could not detect bursts from \sgr was $\sim11$ hours. To obtain a complete picture of the source activity, we searched the data of other satellites that were triggered by the source. The only instrument with a continuous \sgr coverage during the entire day of 2009 January 22 was the {\it INTEGRAL}/SPI-ACS. The SPI-ACS, however, has a higher energy threshold ($>$80~keV) and provides no energy resolution. As a result it detected fewer bursts \citep[slightly over 200][]{mereghetti2009b}, of which only 84 were also detected by the {\it INTEGRAL}/IBIS/ISGRI instrument, and for which spectral information ($>20$ keV) was available \citep{savchenko2010}. Thus the GBM sample is by far the most complete uniform data set for that entire day's activity from \sgrnos.

For each one of the 286 GBM bursts we used the TTE data of those NaI detectors that had a source viewing angle 
(angle between the source direction and the detector normal) of less than 60\degrees. We then checked whether each 
one of these detectors had an unobstructed view of \sgrnos, 
i.e., if no parts of the spacecraft or the Large Area Telescope (LAT) onboard {\it Fermi} were obstructing the view.
This last check is important because several bursts had such a high peak flux that they caused Autonomous Repoint Requests 
of {\it Fermi}, leading to the re-pointing of the spacecraft to the source. The final orientation in 
those cases was 5\degrees~off from the LAT zenith, resulting in the LAT obstructing the view for some detectors. 

\begin{figure}
\begin{center}
\includegraphics[angle=-90,width=0.99\columnwidth]{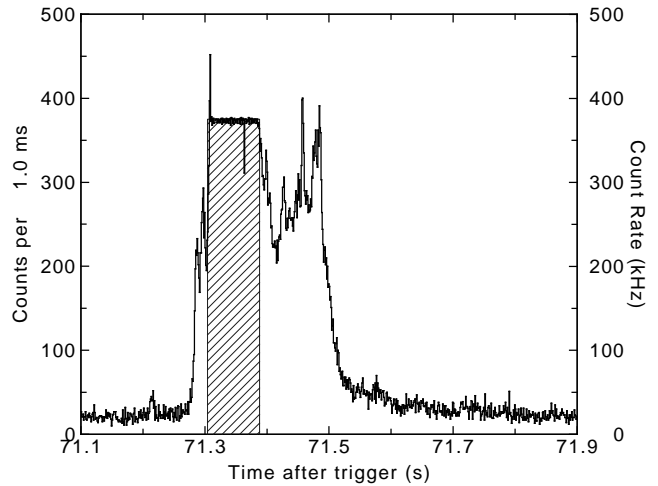}
\caption{One of the 23 saturated \sgr bursts (06:59:34 UT on 22 January 2009). 
The light curve is the sum of the count rates in all 14 detectors across the entire energy range of GBM. 
The saturated time interval is indicated by the hatched area.}
\label{fig:saturation}
\end{center}
\end{figure}

Finally, several bursts in our sample were so bright that they saturated the High Speed Science Data Bus of GBM. This effect occurs
when the total TTE count rate of all detectors exceeds a limit of 375,000 counts s$^{-1}$. 
Out of the 286 bursts only 23 were affected; an example of a saturated burst is shown in Figure~\ref{fig:saturation} (hatched area). 
We did not use the saturated bursts in our temporal analyses (Section~\ref{sec:temp}), 
since the durations, emission times and peak times of these bursts are severely affected. 
We performed a spectral analysis using the non-saturated parts of these bursts, 
and used their fluences only in the compilation of the cumulative energy fluence emitted during the active period of \sgr (Section~\ref{sec:spec}). 
None of the 23 saturated bursts were used in any of the other parts of the spectral analysis and discussion sections in this paper.

\section{Temporal analysis}\label{sec:temp}

\begin{figure*}
\begin{center}
\includegraphics[angle=-90,width=0.270\textwidth]{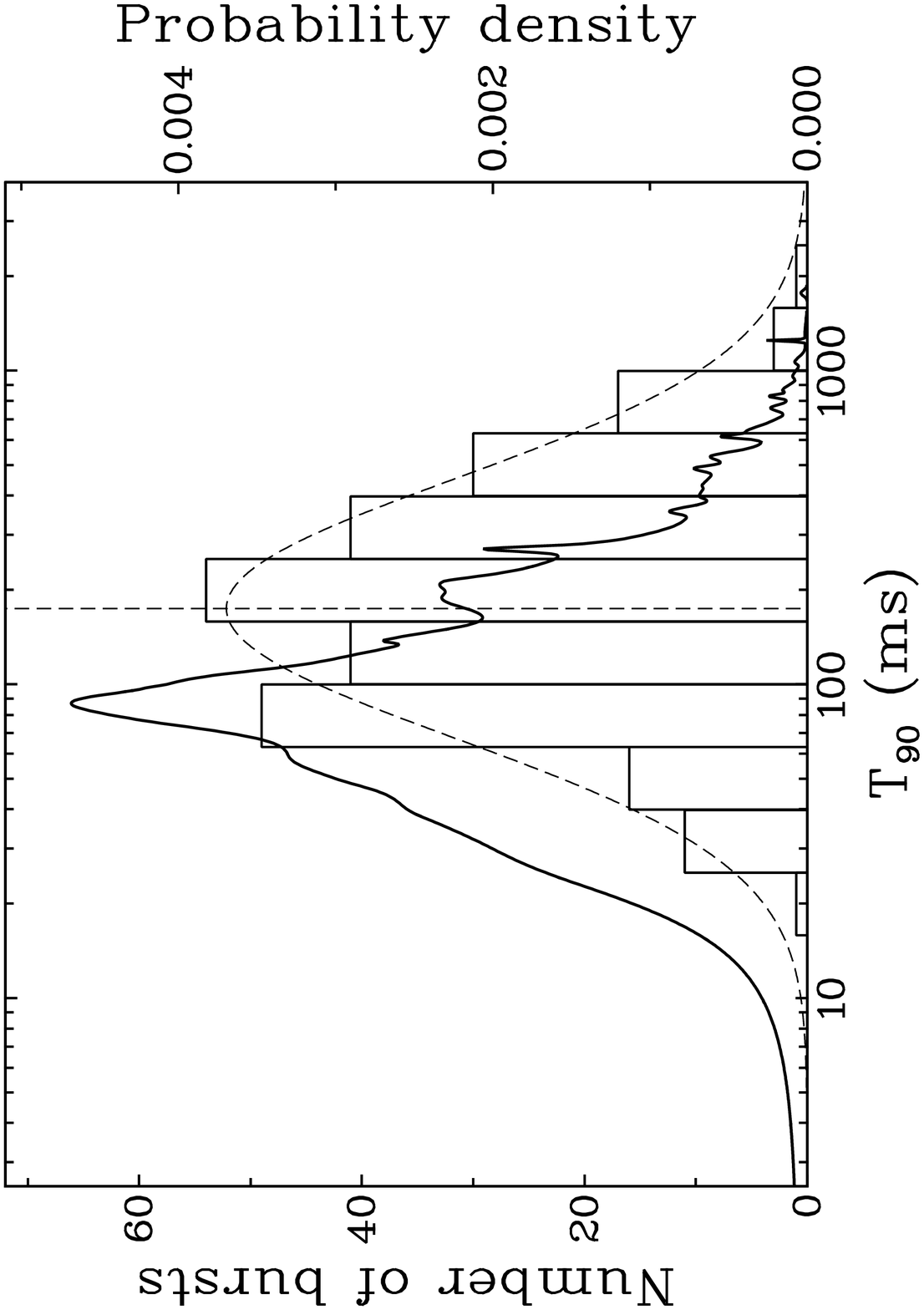} 
\includegraphics[angle=-90,width=0.235\textwidth]{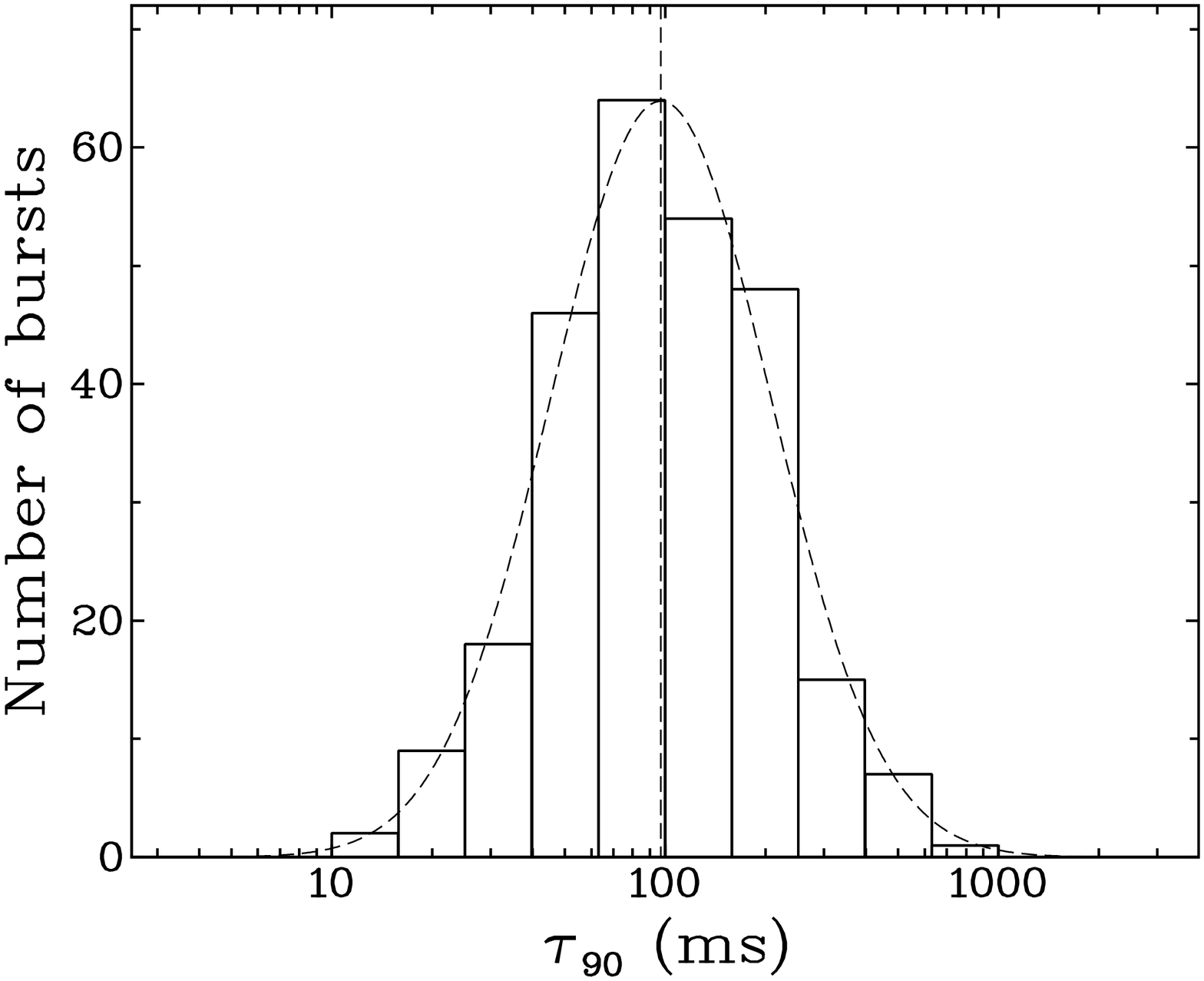} 
\includegraphics[angle=-90,width=0.235\textwidth]{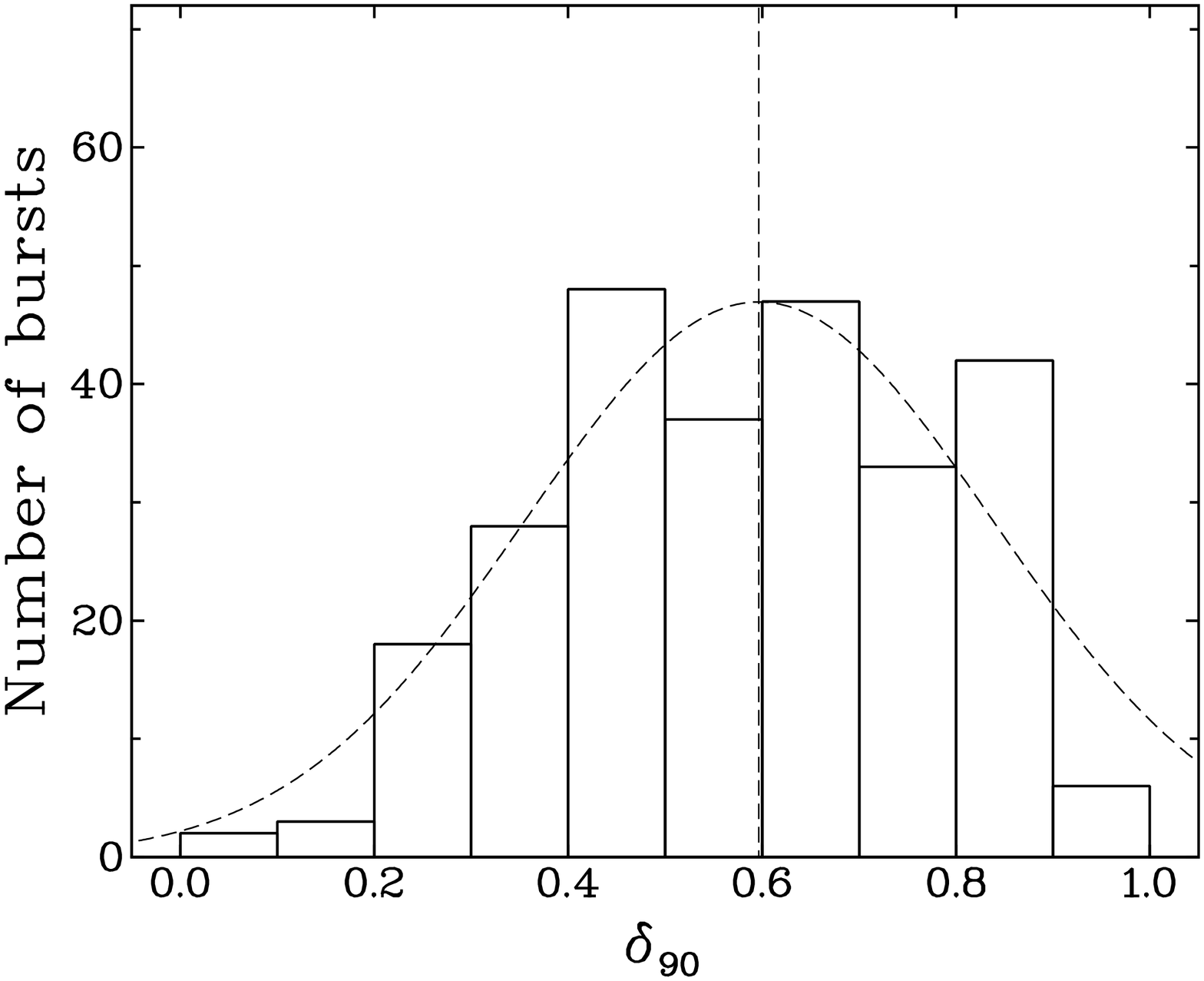} 
\includegraphics[angle=-90,width=0.235\textwidth]{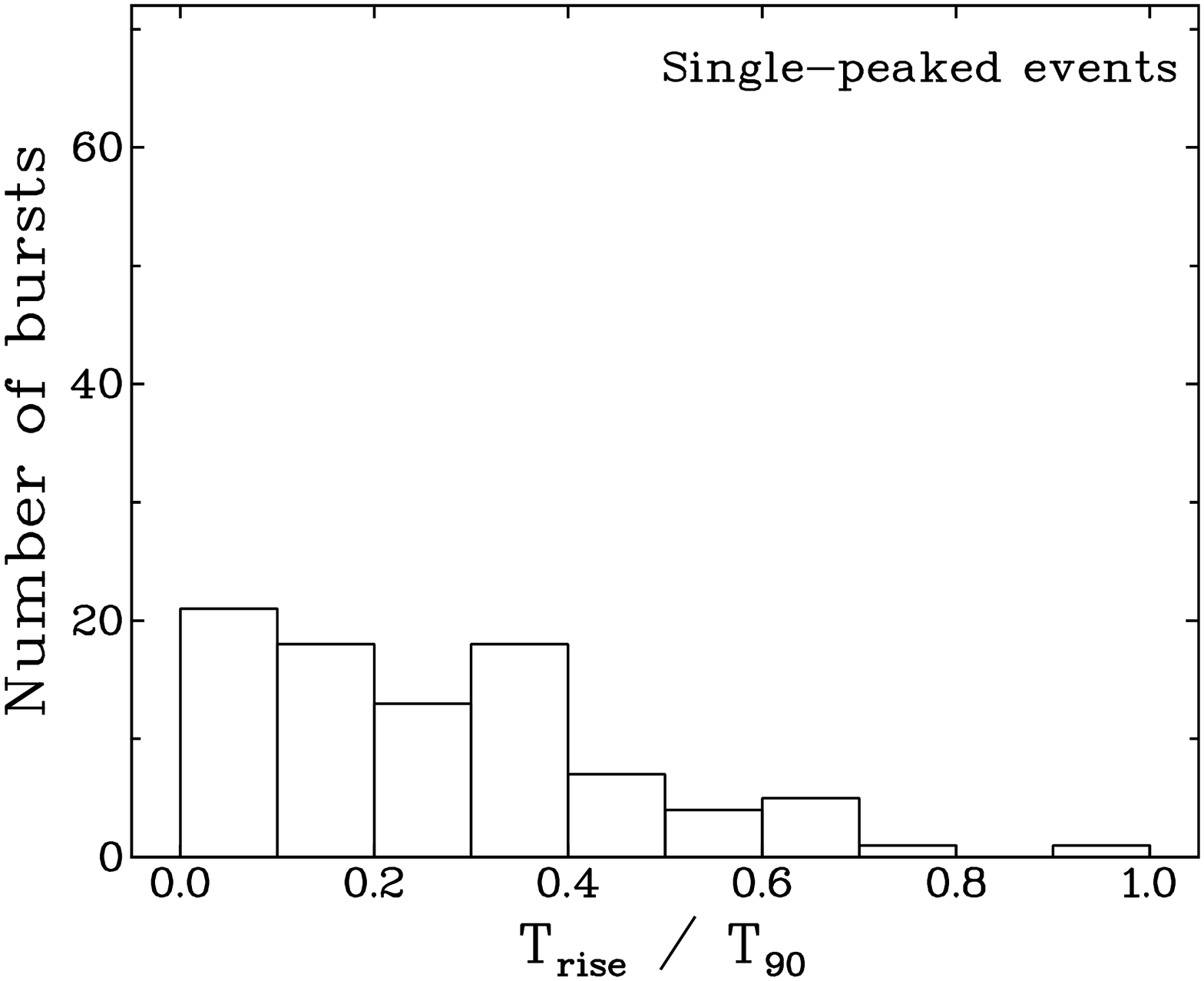} \\ \vspace{0.1cm}
\includegraphics[angle=-90,width=0.270\textwidth]{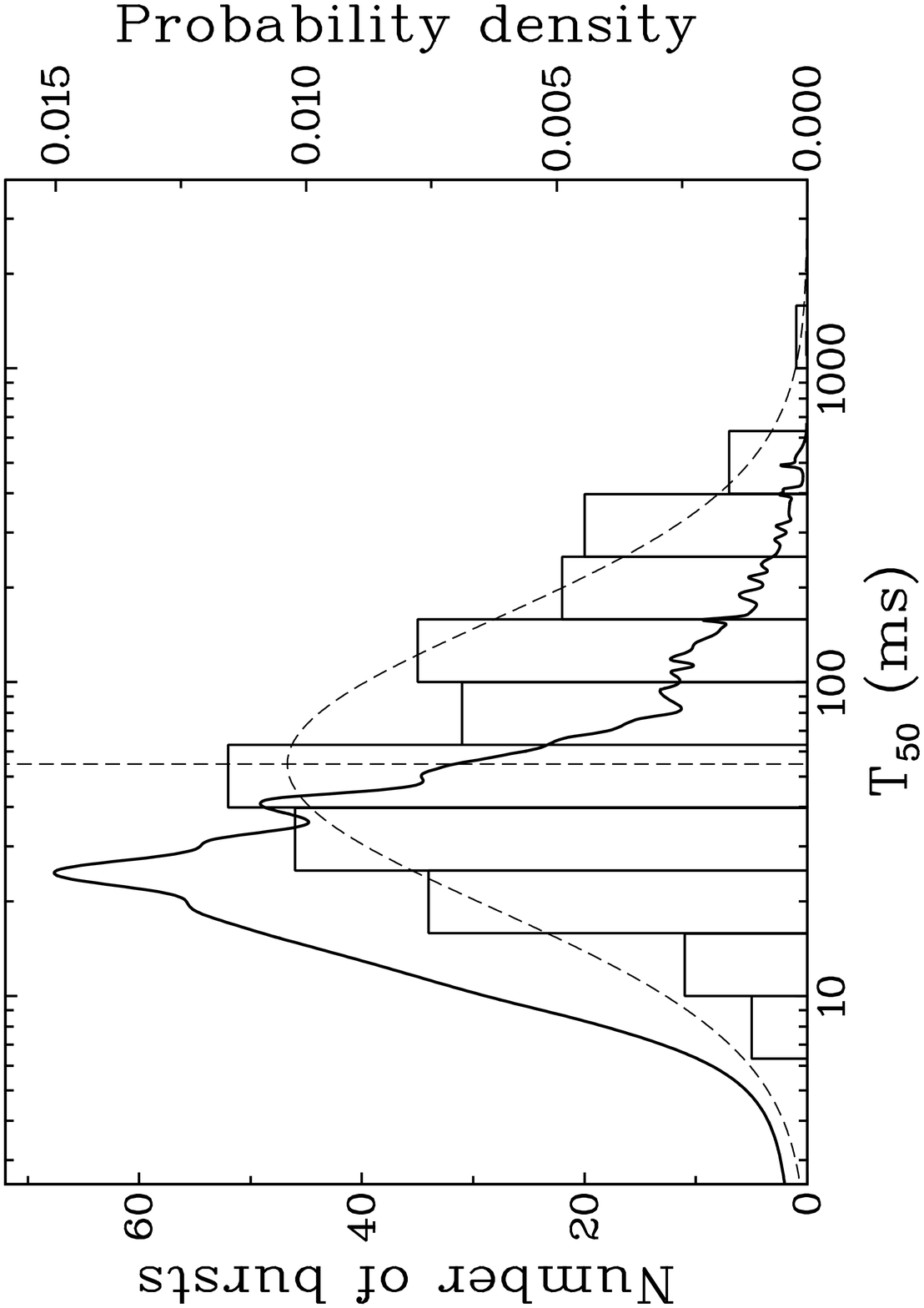} 
\includegraphics[angle=-90,width=0.235\textwidth]{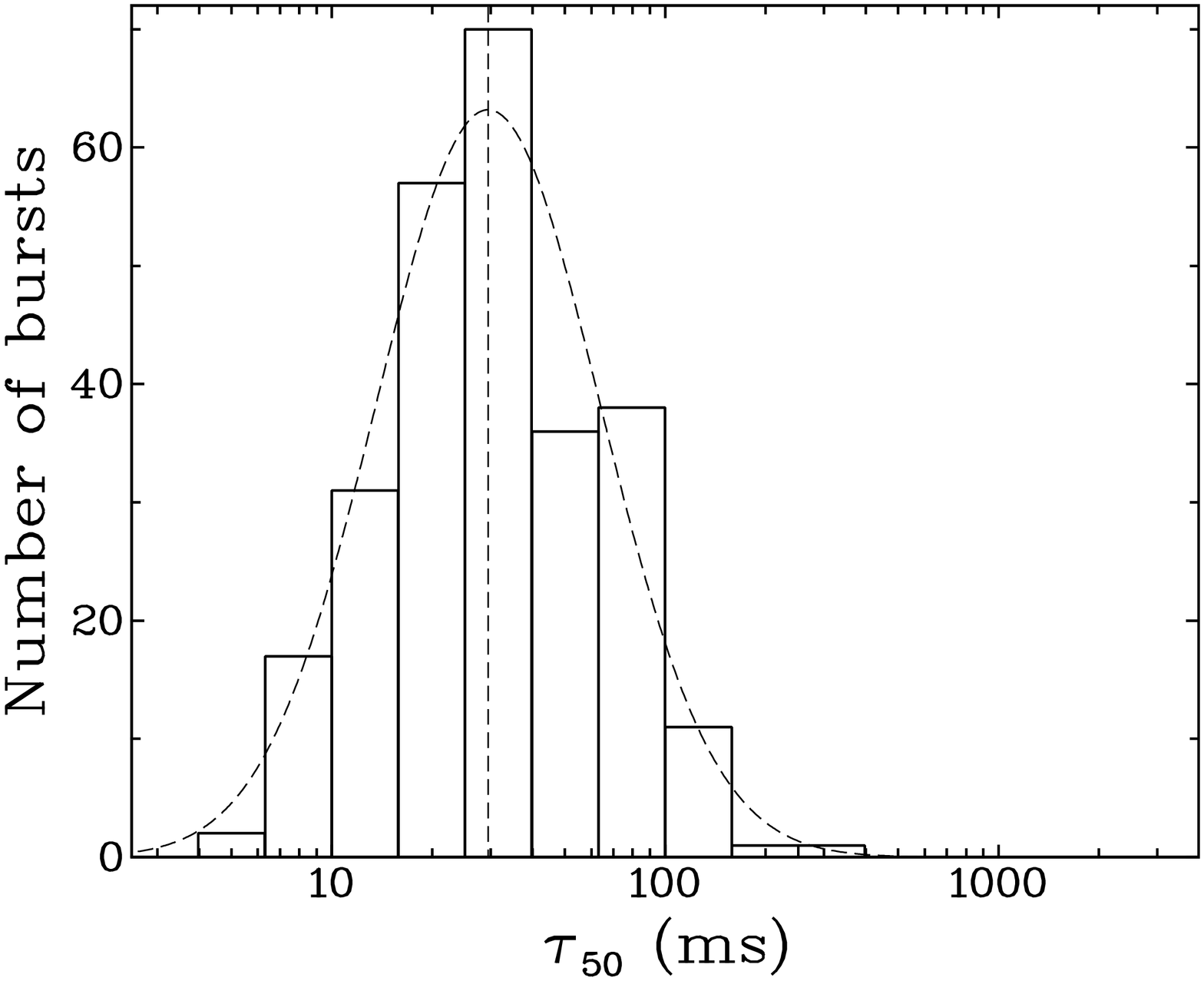} 
\includegraphics[angle=-90,width=0.235\textwidth]{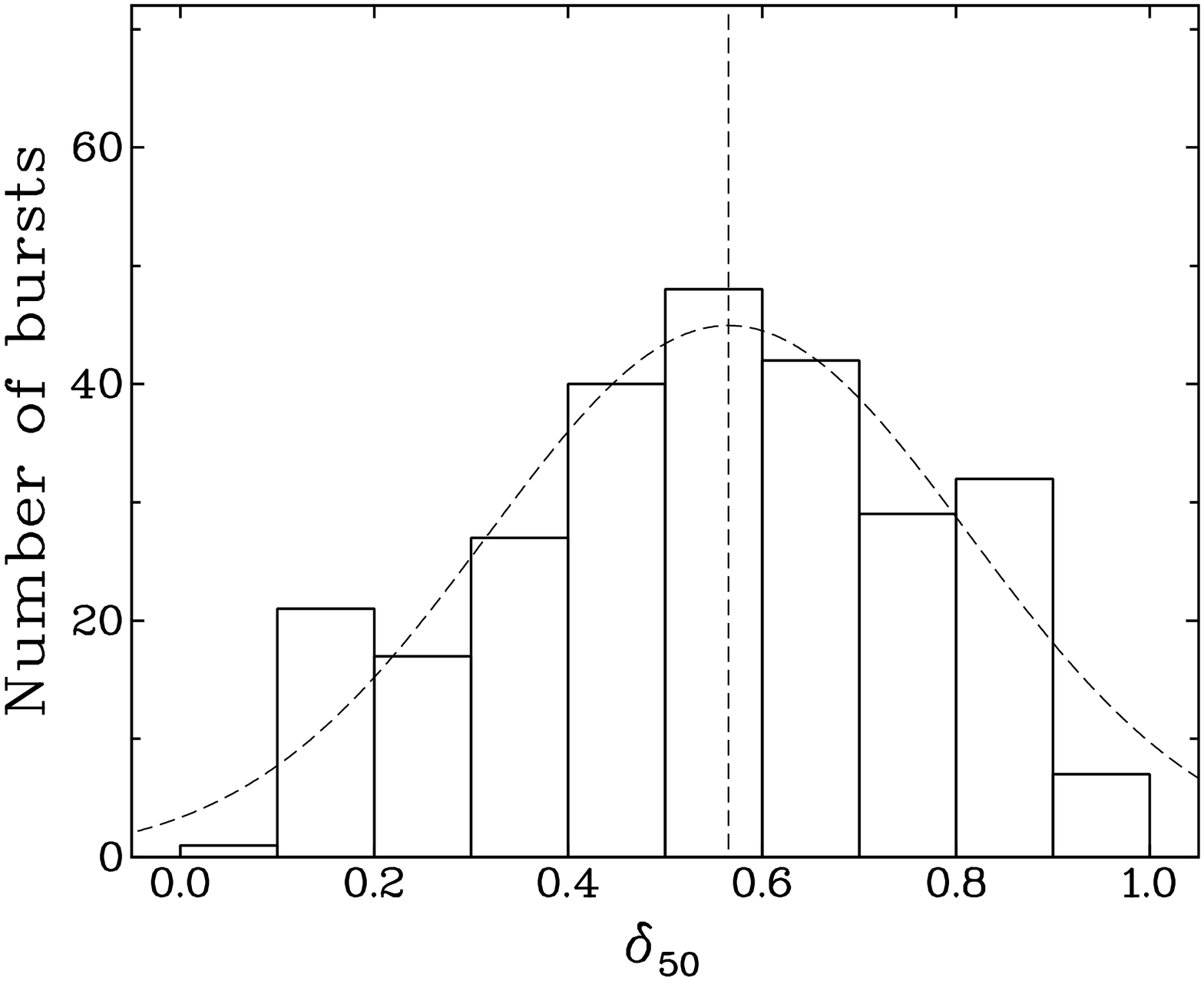} 
\includegraphics[angle=-90,width=0.235\textwidth]{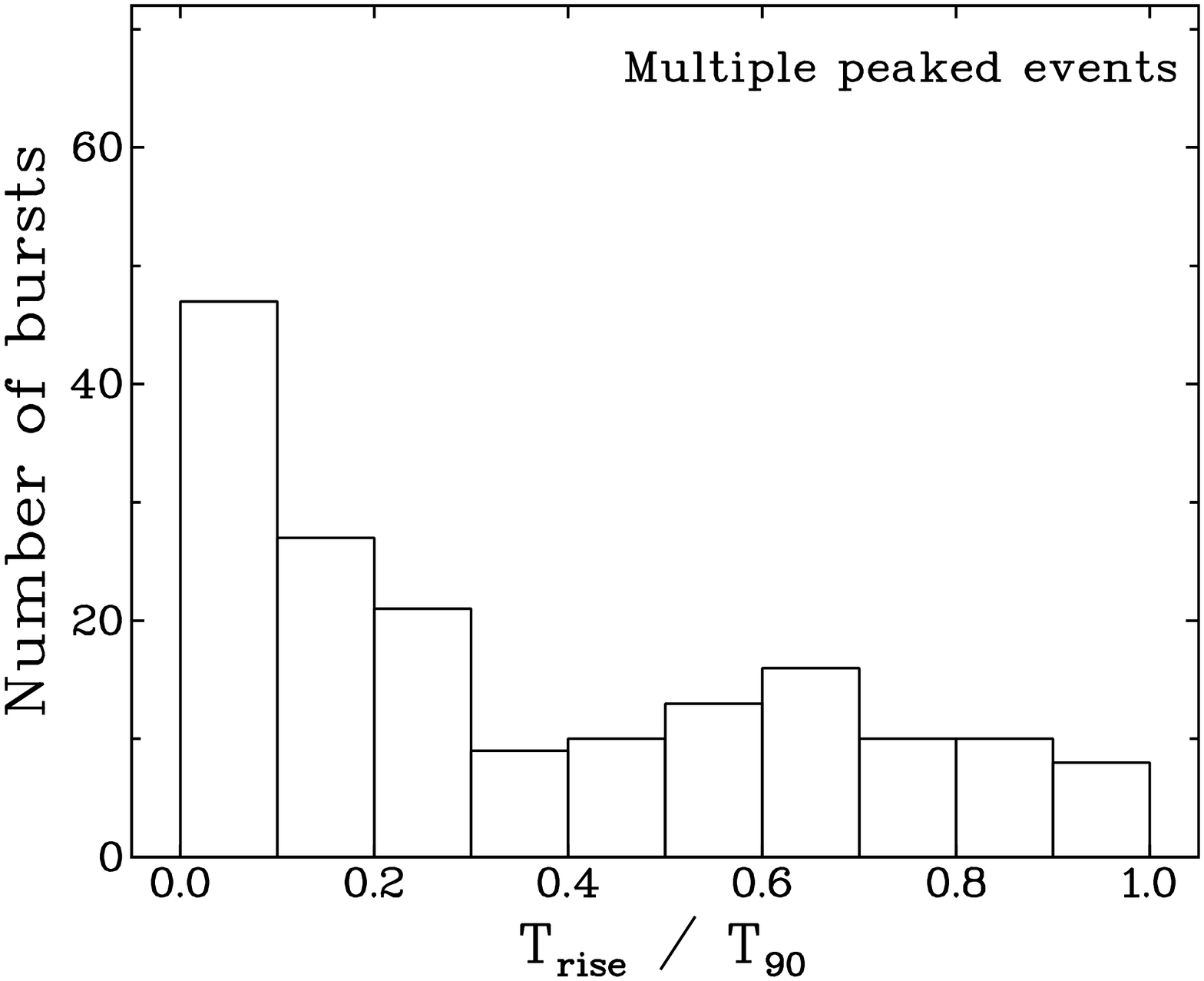}
\caption{Distributions of the durations T$_{90,50}$ (first column panels); 
emission times $\tau_{90,50}$ (second column panels); 
duty cycles $\delta_{90,50}$ (third column panels); and T$_{\rm{rise}}$/T$_{90}$ (fourth column panels) 
for single-peaked (top) and multiple peaked (bottom) events. 
The dashed lines show the best-fit log-normal (T$_{90,50}$ and $\tau_{90,50}$) 
and normal ($\delta_{90,50}$) functions, with the vertical dashed lines indicating the fitted mean values. 
For T$_{90,50}$ the probability density functions are shown as solid lines.}
\label{fig:temp}
\end{center}
\end{figure*}

To characterize the temporal properties of all the bursts in our sample, 
we determined their durations, emission times, duty cycles and rise times. 
We calculate these quantities using TTE data binned at 2~ms resolution in the $8-100$~keV energy range. 
We apply a two-step background subtraction method to the total number of counts during and around the burst time interval; 
details on the particular method used here are given in \citet{gogus2001} and \citet{lin2011}. 

The T$_{90}$ (T$_{50}$) duration is the time during which 90\% (50\%) of the total burst counts are accumulated \citep{kouveliotou1993}. 
Another parameter to quantify burst durations is the emission time $\tau_{90}$ ($\tau_{50}$) \citep{mitrofanov1999}. 
In contrast with T$_{90,50}$, the emission time does not necessarily span a time interval of consecutive time bins; it is computed by adding time bins ordered from the highest to the lowest number of counts until 90\% or 50\% of the burst counts is accumulated.  
We then calculated for each burst the duty cycle $\delta_{90}$ ($\delta_{50}$), 
which is the ratio of $\tau_{90}$ ($\tau_{50}$) over T$_{90}$ (T$_{50}$); this quantity gives an indication 
of how sharply peaked a given burst light curve is \citep{mitrofanov1999}. 
The final temporal parameter we determined is the burst rise time T$_{\rm{rise}}$ (the time for a burst 
to reach its peak count rate); in multiple peaked events, we used the highest count rate bin as the peak.

\begin{table}
\begin{center}
\caption{Temporal analysis of 263 \sgr bursts}
\label{tab:temp}
\renewcommand{\arraystretch}{1.25}
\begin{tabular}{|l|l|l|l|}
\tableline
Parameter & Fit mean\footnote{log-normal fit for T$_{90,50}$ and $\tau_{90,50}$, normal fit for $\delta_{90,50}$} & $\sigma$\footnote{in the log-frame except for $\delta_{90,50}$} & Mean \\
\tableline
T$_{90}$ (ms) & $174\pm10$ & $0.41\pm0.02$ & 258 \\
T$_{50}$ (ms) & $55\pm4$ & $0.46\pm0.03$ & 104 \\
$\tau_{90}$ (ms) & $97\pm3$ & $0.33\pm0.01$ & 127 \\
$\tau_{50}$ (ms) & $29\pm2$ & $0.34\pm0.02$ & 39 \\
$\delta_{90}$ & $0.60\pm0.03$ & $0.24\pm0.04$ & 0.58 \\
$\delta_{50}$ & $0.56\pm0.03$ & $0.25\pm0.03$ & 0.54 \\
\tableline
\end{tabular}
\end{center}
\end{table}

The results of our temporal analysis are shown in Figure~\ref{fig:temp} and Table~\ref{tab:temp}. 
Figure~\ref{fig:temp} shows the distributions of T$_{90,50}$ (first column panels), $\tau_{90,50}$ (second column panels) and  $\delta_{90,50}$ (third column panels), 
and T$_{\rm{rise}}$/T$_{90}$ for single-peaked and multiple peaked events (fourth column panels). 
We fit the distributions of the durations and emission times with log-normal functions, 
and the duty cycle distributions with normal functions. 
The means and standard deviations of these functions are given in Table~\ref{tab:temp}, 
together with their normal mean values. 

Next we used the T$_{90,50}$ values and their uncertainties to construct their probability density functions (PDFs) as described in
\citet{lin2011}. 
For each individual PDF we adopted a two-sided normal distribution with the widths given by the measured (asymmetric) uncertainties \citep{starling2008}. 
Figure~\ref{fig:temp} shows the average of all the individual PDFs for T$_{90,50}$. 
Note that the PDFs peak at smaller values than the duration histograms, namely at $\sim85$~ms and $\sim25$~ms for T$_{90}$
and T$_{50}$, respectively. 
This is due to the fact that the shorter (and weaker) events have relatively larger asymmetries in their uncertainties. 

The histograms shown in Figure~\ref{fig:temp} for the various temporal parameters are typical of magnetar bursts,
although there may be variations between different sources \citep[e.g.,][]{gogus2001,gavriil2004,lin2011}. 
The mean T$_{90}$ values in the GBM energy band typically range between 0.1-0.2\,s, while $\tau_{90}$ ranges are 0.05-0.1\,s, resulting in duty cycles of $0.4-0.6$ \citep[see also][]{lin2011}. 
Our mean T$_{90}$ values for \sgr are larger than those found for {\it INTEGRAL} \citep[68~ms;][]{savchenko2010}, but smaller than the {\it Swift} bursts \citep[305~ms;][]{scholz2011}. 
This is due to the different sensitivities and energy ranges of the instruments \citep[see also][]{scholz2011}. 
Regarding the rise times, we confirm what has been established for other magnetar bursts \citep[e.g.,][]{gogus2001,scholz2011}: 
the asymmetric T$_{\rm{rise}}$/T$_{90}$ distributions for single-peaked events indicate 
that magnetar bursts have shorter rise than decay times, 
and the bimodal distribution of multiple peaked events shows that for most bursts the first peak is the brightest. 
In Section~\ref{sec:disc} we elaborate further on the correlation of the temporal and spectral properties of \sgrnos, and compare these to similar correlations in other magnetar sources.

\section{Spectral Analysis}\label{sec:spec}

\begin{figure*}
\begin{center}
\includegraphics[angle=90,width=0.33\textwidth,trim=15 60 25 -15,clip=true]{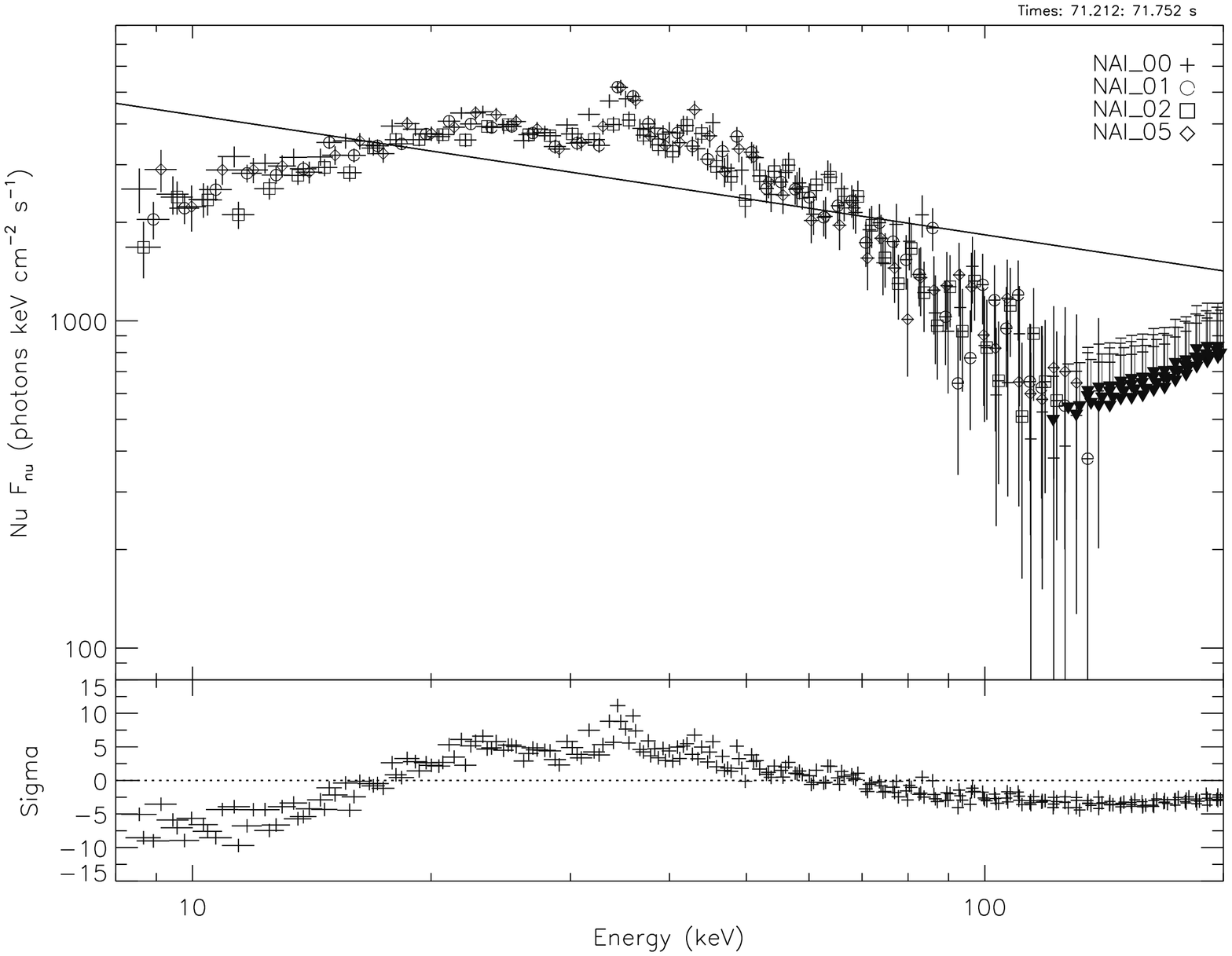}
\includegraphics[angle=90,width=0.33\textwidth,trim=15 60 25 -15,clip=true]{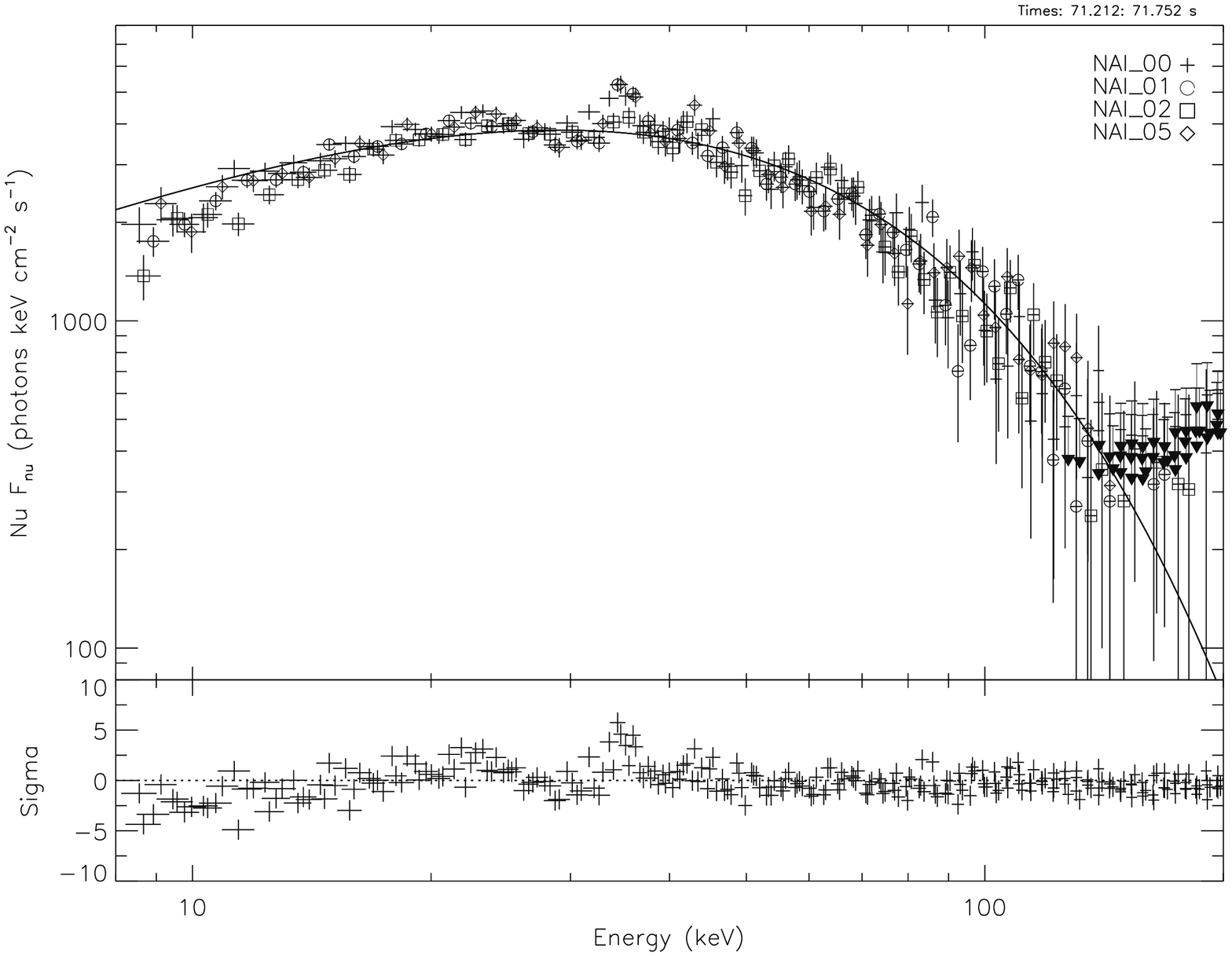}
\includegraphics[angle=90,width=0.33\textwidth,trim=15 60 25 -15,clip=true]{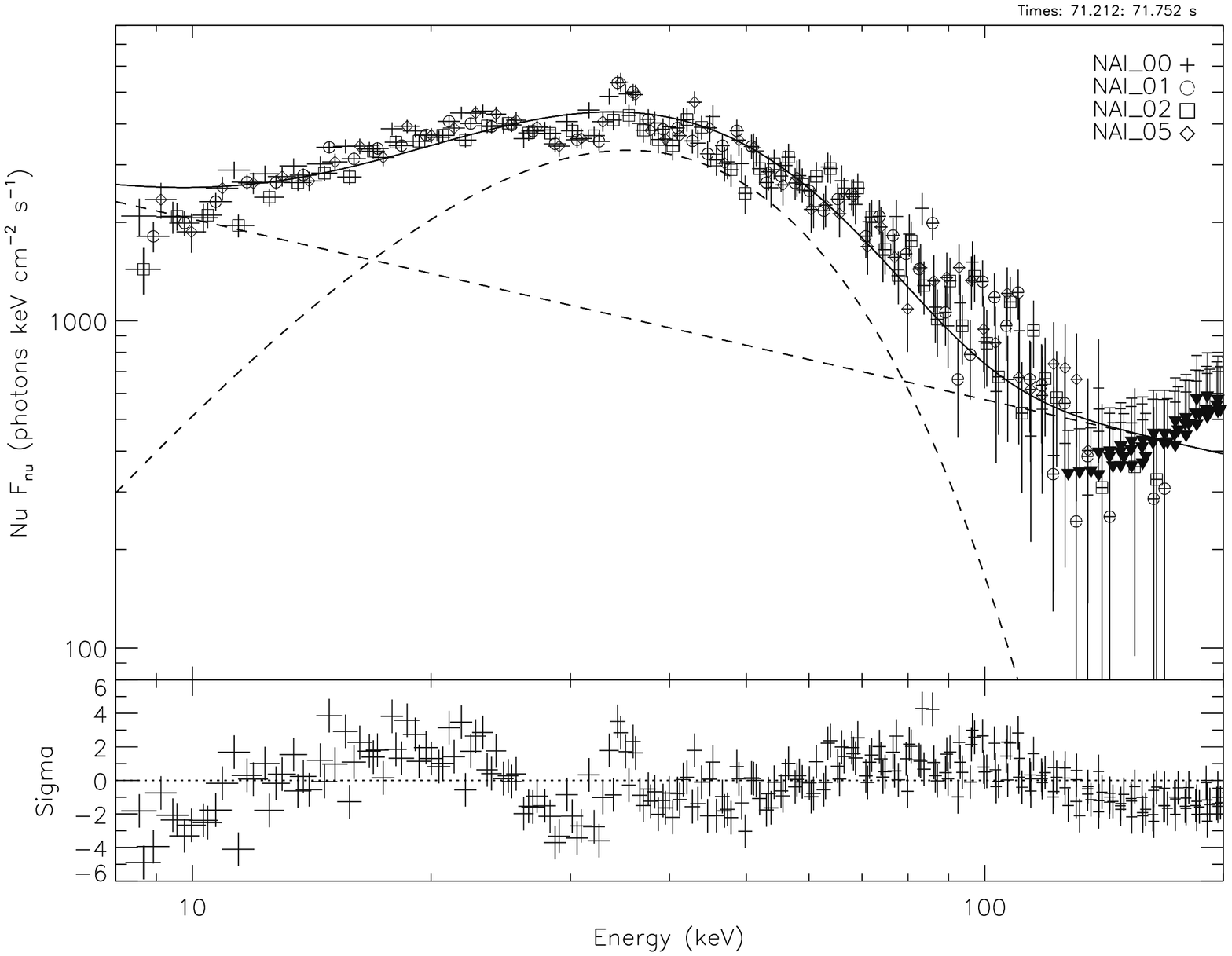} \\ \vspace{0.1cm}
\includegraphics[angle=90,width=0.33\textwidth,trim=15 60 25 -15,clip=true]{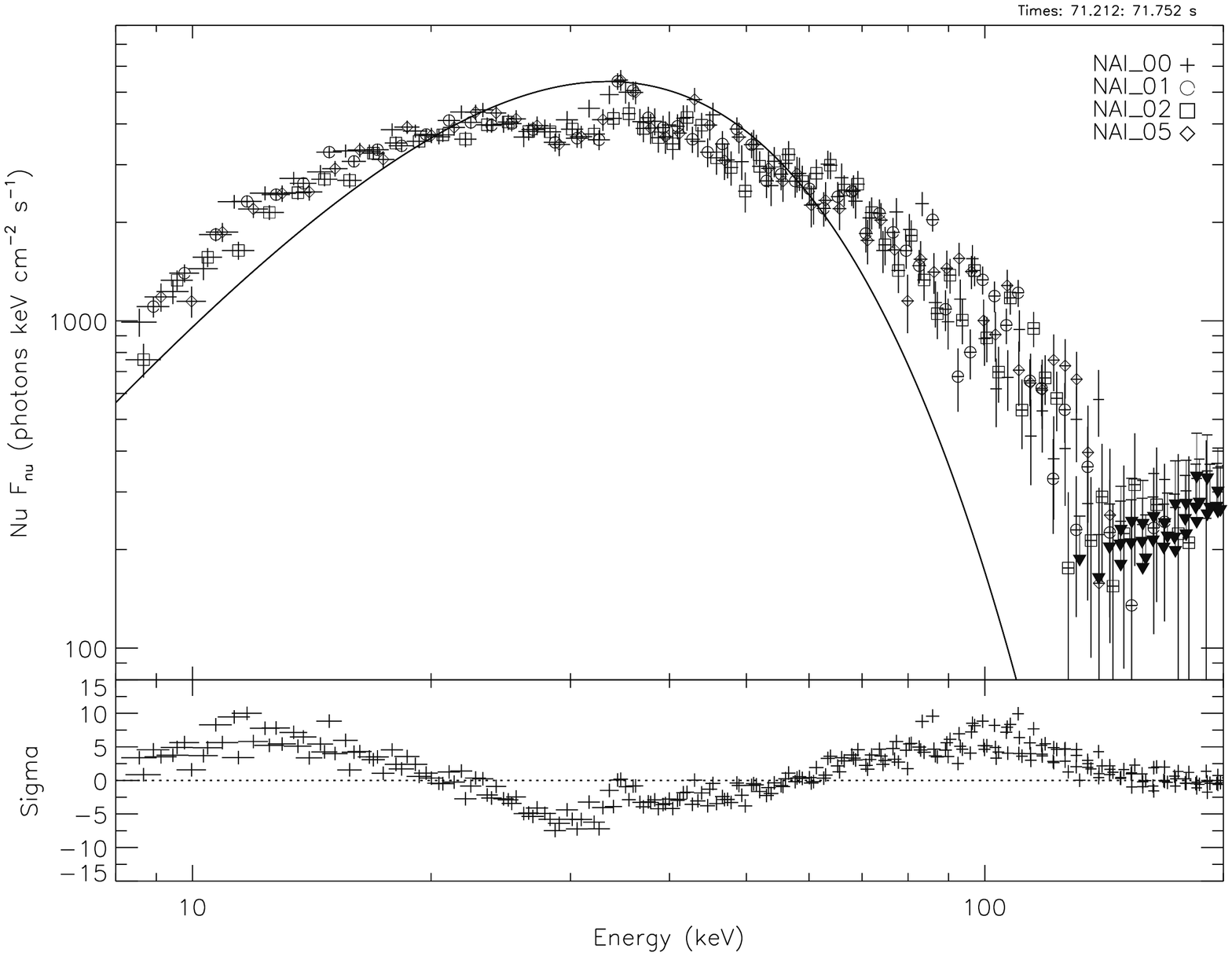}
\includegraphics[angle=90,width=0.33\textwidth,trim=15 60 25 -15,clip=true]{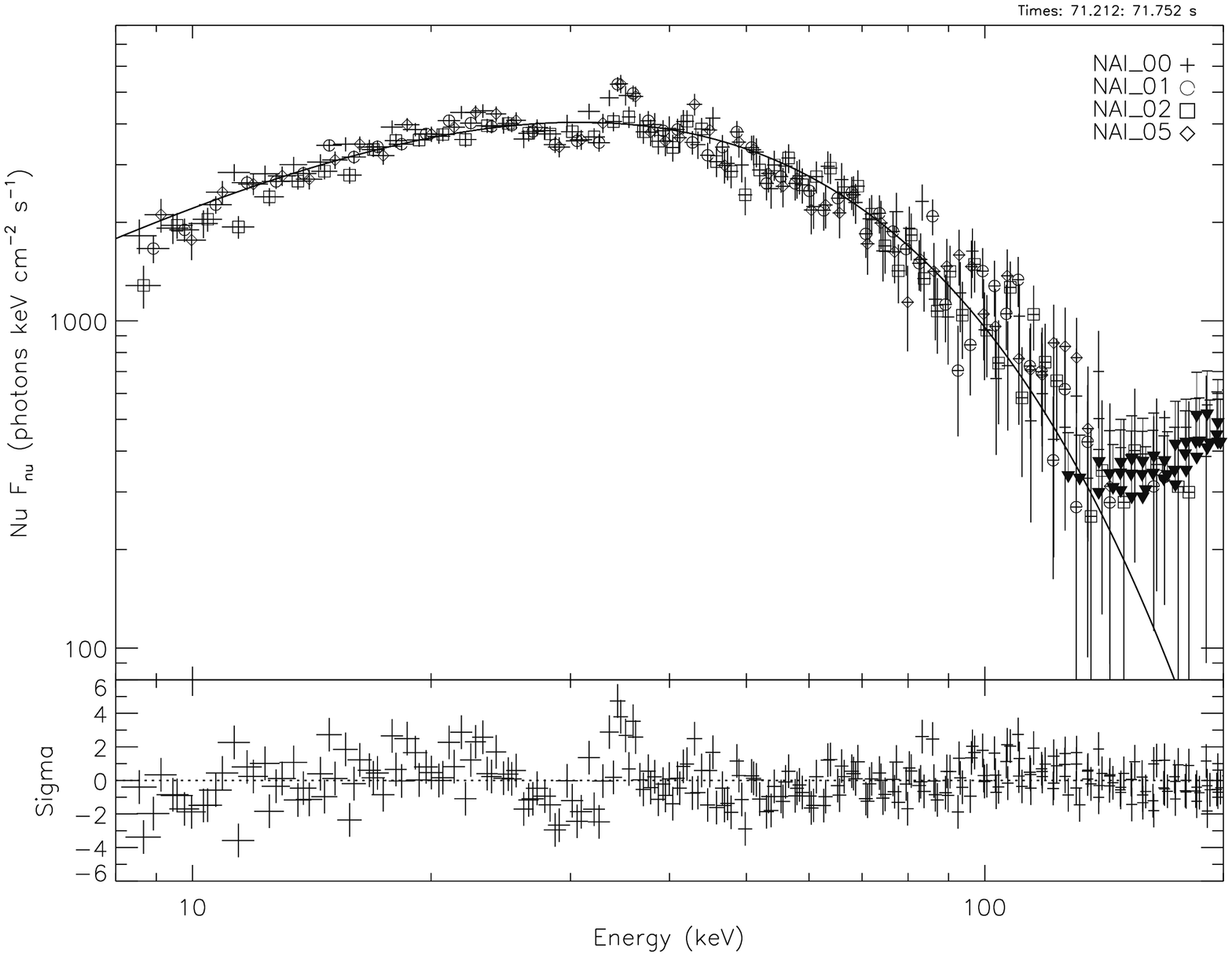}
\includegraphics[angle=90,width=0.33\textwidth,trim=15 60 25 -15,clip=true]{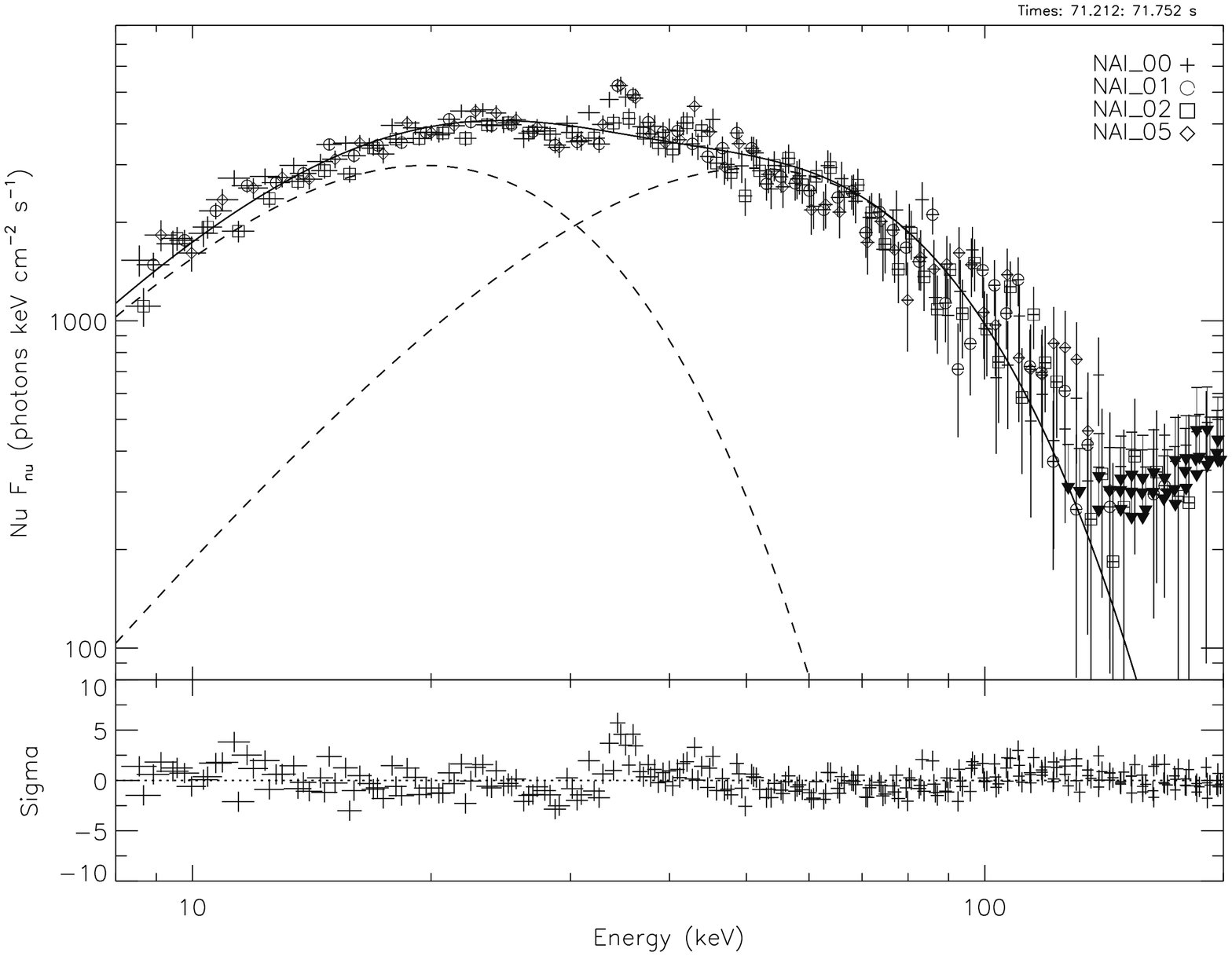}
\caption{Time-integrated $\nu$F$_{\nu}$ spectra of one of the brightest \sgr bursts (06:59:34 UT on 22 January 2009), 
shown in Figure~\ref{fig:saturation}, for various spectral models: 
power law (top left), single black body (bottom left), OTTB (top center), 
Comptonized (bottom center), power law plus black body (top right), 
and two black body functions (bottom right). 
The bottom panel of each spectrum shows the fit residuals. 
The feature between 30 and 40~keV in the fit residuals is a result of the imperfect modeling of the NaI K-edge.}
\label{fig:spectra}
\end{center}
\end{figure*}

We performed time-integrated spectral analysis of all 286 bursts in our sample using the spectral analysis software package {\it RMFIT (3.3rc8)}\footnotemark{},
\footnotetext{http://fermi.gsfc.nasa.gov/ssc/data/analysis/user/}
which was developed specifically for GBM data analysis. 
For every burst we selected intervals of background without bursts present 
before and after a burst, and fit them with a polynomial
of the third or fourth order. 
For the spectral fits we generated Detector Response Matrices using {\it GBMRSP v1.81}. 
To obtain the best fit parameters 
for any given spectral model we minimized the Castor C-statistic (C-statistic hereafter). This method is used to fit data with a low number of counts and is a modification\footnotemark{}
\footnotetext{heasarc.gsfc.nasa.gov/docs/xanadu/xspec/wstat.ps} of the Cash statistic so that it asymptotically distributes as $\chi^2$.  

We fit various spectral models to the TTE data of each burst: 
a power law (PL), a black-body function (BB), optically thin thermal bremsstrahlung (OTTB), 
a power law with an exponential cutoff (Comptonized model), 
a combination of a power law and a blackbody function (PL+BB), 
and two black-body functions (BB+BB). 
Figure~\ref{fig:spectra} shows spectra of one of our brightest bursts (using only the unsaturated parts), whose light curve is shown in Figure~\ref{fig:saturation}. 
We have used in this spectral analysis counts from 
four NaI detectors following the detector selection criteria given in Section~\ref{sec:sample}. 
The $\nu$F$_{\nu}$ spectra in Figure~\ref{fig:spectra} show fits with the six different spectral models 
mentioned above and their fit residuals. 
The fit residual trends shown here are very similar for all bursts in our sample, 
with the brightest bursts presenting the strongest trends. 
From Figure~\ref{fig:spectra} an obvious conclusion would be that OTTB, Comptonized and BB+BB fits
provide a better description of the data than PL, BB or PL+BB. 
Section~\ref{sec:sims} attempts to quantify this statement. 
We note that the apparent feature between 30 and 40~keV in the fit residuals 
is a result of the NaI K-edge, which has not been modeled perfectly in the GBM calibration \citep{bissaldi2009}. 
We have performed fits with and without the K-edge energy channels, 
and found that their inclusion does not change any of the fit parameters significantly, therefore we included these channels in all fits for better statistics.

\subsection{Simulations}
\label{sec:sims}

\begin{figure*}
\begin{center}
\includegraphics[angle=-90,width=0.32\textwidth]{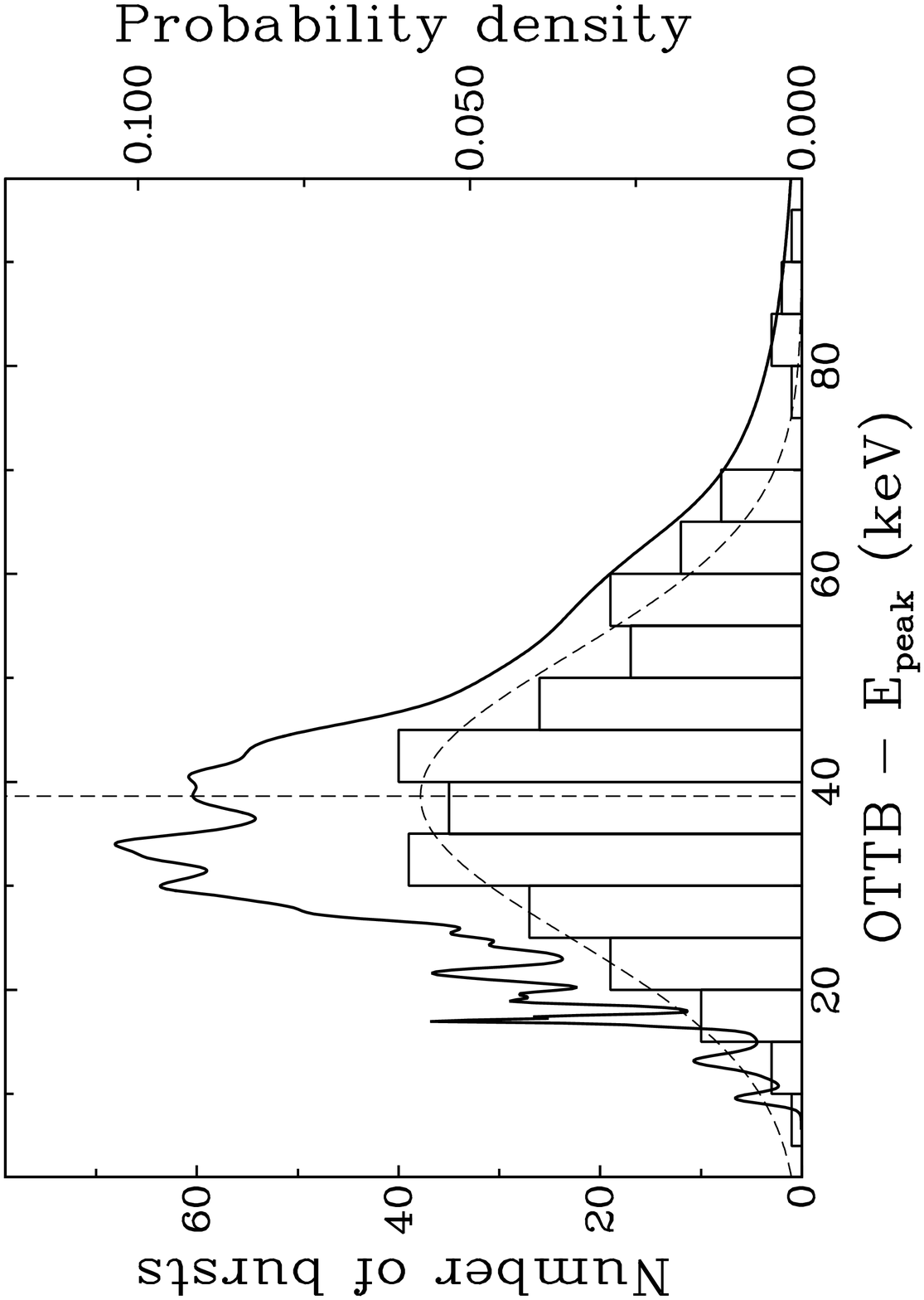} \hspace{0.15cm}
\includegraphics[angle=-90,width=0.32\textwidth]{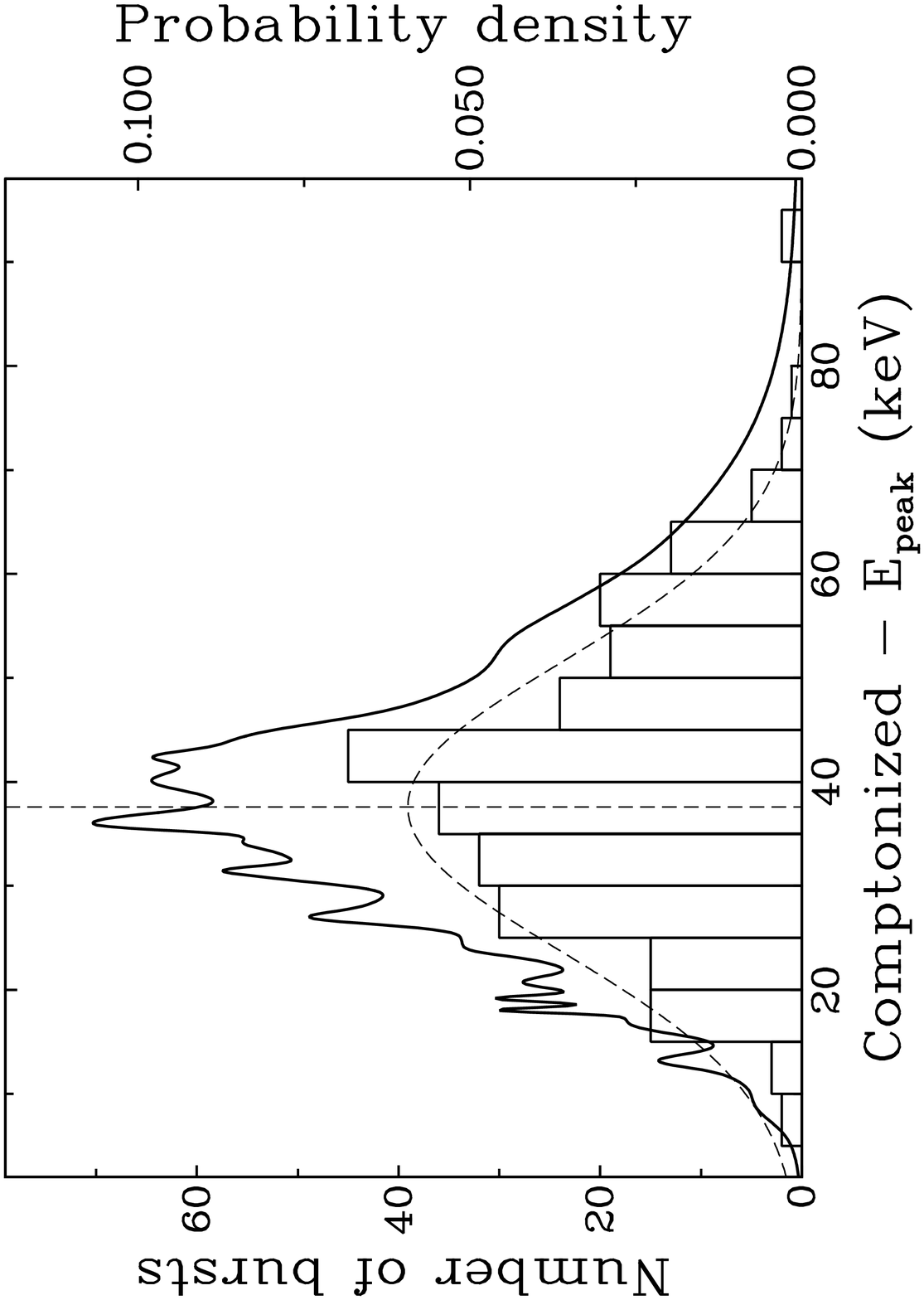} \hspace{0.15cm} 
\includegraphics[angle=-90,width=0.32\textwidth]{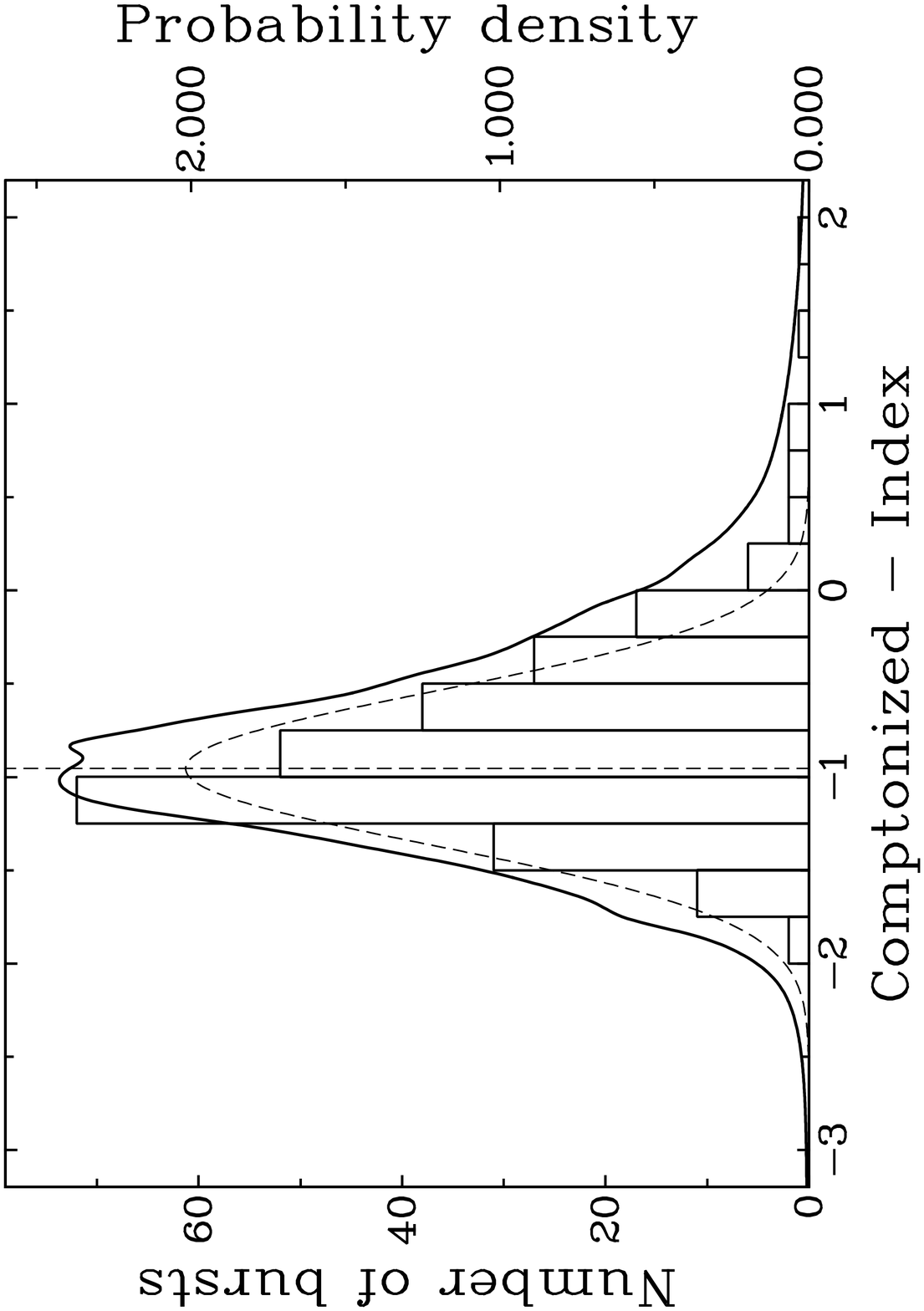} 
\caption{Distributions of the OTTB $E_{\rm{peak}}$ (left), Comptonized $E_{\rm{peak}}$ (middle), 
and of the Comptonized power-law index (right). 
The dashed lines show the best-fit normal functions, with the vertical dashed lines indicating the fitted mean values. The
probability density functions are shown as solid lines.}
\label{fig:ottbcomp}
\end{center}
\end{figure*}

We have performed extensive simulations with {\it RMFIT} (3.4rc3) to evaluate the effect of statistical fluctuations on identifying the best model to describe SGR burst spectra. 
For this purpose we selected six bright bursts \citep[two from \sgr and three from SGR\,J0501$+$4516; see also][]{lin2011}, and compared pairs of various models, generating 30,000 synthetic spectra for each burst and each relevant detector. 
Each synthetic spectrum was based on the sum of the predicted source counts and the measured background counts in each energy channel. 
The former counts were computed from the analytical function with the largest C-statistic used to fit the real data (null hypothesis) folded with the instrumental response function of the relevant detector (as a check, we also used as null hypothesis the function giving the lowest C-statistic and we obtained consistent results). 
The background counts were estimated for each detector from the real data. 
Next, we applied Poisson fluctuations to the summed counts to obtain the final synthetic spectrum.

During the fit process we generated a synthetic background spectrum by adding Poisson fluctuations to the real background estimate in each energy channel, and subtracting it from the synthetic spectrum. 
The resulting spectrum was fit with the two models of each pair. 
We checked and confirmed that the input spectral parameters were well recovered in the fits of the synthetic spectra, by comparing them and their statistical errors to the simulated parameter distributions and their uncertainties. 
Finally, to test whether this recovery was intensity dependent, we also simulated two weak events and arrived at the same conclusions.

We then proceeded to compare the fits between different model pairs. 
To this end we computed the C-statistic difference between one input model and each other spectral model for all 30,000 synthetic spectra per burst, and compared the resulting distribution to the difference obtained using the same models with the real burst data.

First we compared PL and BB with all other spectral models, because the former systematically gave the largest fit residual patterns for the majority of the events in our sample. 
For all simulated events, we found that statistical fluctuations in the signal and background cannot explain that OTTB or any of the more complex models fit the data better than PL or BB, including the weakest one with a fluence of $6.7\times10^{-8}$~erg/cm$^2$. 
Using the same procedure, we then compared different models, namely OTTB, Comptonized, PL+BB and BB+BB. 
Our simulations showed that the statistical fluctuations in the data could account for the difference in C-statistic values between these models, preventing us from drawing conclusions on the best spectral shape. 
We cross-checked and confirmed the results of our simulations using XSPEC \citep[v12.6;][]{arnaud1996}. 
We conclude, therefore, that we cannot unambiguously determine whether OTTB, Comptonized, PL+BB or BB+BB is the best spectral fit model for our burst sample.

The same method has been used to distinguish between spectral models for gamma-ray bursts (GRBs) detected with GBM \citep[see e.g.][]{guiriec2011}, albeit resulting in identification of the best-fit spectral model. 
The discrepancy with the SGR simulation results here could be explained by the fact that even the brightest SGR bursts have a lower total number of source counts mostly distributed over a much narrower energy range (10$-$50~keV).

\subsection{Spectral Model Selection}

Although our simulation results were inconclusive regarding the preferred spectral model, we attempt here to narrow down the number of models that give a good description of the SGR burst spectra. 
We did this down-selection based on the following arguments. 
When fitting the spectra of individual bursts, we obtained systematically large fit residual patterns with the PL, BB, and PL+BB functions for a large fraction of the events in our sample; all other model residuals were randomly distributed. 
For the PL and BB models we showed in Section  \ref{sec:sims} that they give a worse description of the data than all other models for bursts with fluences above $6.7\times10^{-8}$~erg/cm$^2$. 
In addition, the PL and BB models had predominantly the largest C-statistic values for all bursts compared to all other models, resulting in very large C-statistic differences, typically tens to hundreds and for some events even a few thousand. 
This difference increased with fluence, indicating that the brightest events are worst fit with a PL or BB. 
Such a dependence on brightness was not seen when comparing C-statistic differences between all other models. 
The combination of all these arguments led us to discard PL and BB models from further analysis for bursts at all fluences. 
Our PL+BB fits gave a C-statistic value better than OTTB for only one third of the bursts in our sample, while PL+BB has two more free parameters than OTTB. Moreover, the PL+BB model systematically gave residual patterns not seen in the BB+BB, Comptonized and OTTB fits. 
We, therefore, excluded the former model from further consideration. 

It is not possible to exclude any of the three remaining models, namely OTTB, Comptonized and BB+BB. 
We cannot choose between OTTB and Comptonized because the power-law index we retrieve in the Comptonized fits is often very close to $-1$ (in photon space) and OTTB is basically a special case of Comptonized with an index of $-1$. 
This is illustrated in the rightmost panel of Figure~\ref{fig:ottbcomp}, which shows the distribution of the Comptonized power-law indices for all events, together with a normal fit to the distribution with a mean value of $-0.95$ and a width of 0.41. Finally, BB+BB is very similar to the Comptonized function over this relatively small energy range (e.g., see the middle right and bottom right panels of Figure~\ref{fig:spectra}). 
When comparing the C-statistic for individual bursts, the difference between these two models lies almost always between $-25$ and 25, with the majority between $-10$ and 10, making the choice of the best fit very difficult. We, therefore, describe below our results of the spectral fits on all 263 events using the OTTB, Comptonized and BB+BB models.

\subsection{OTTB and Comptonized Model}

\begin{figure*}
\begin{center}
\includegraphics[angle=-90,width=0.33\textwidth]{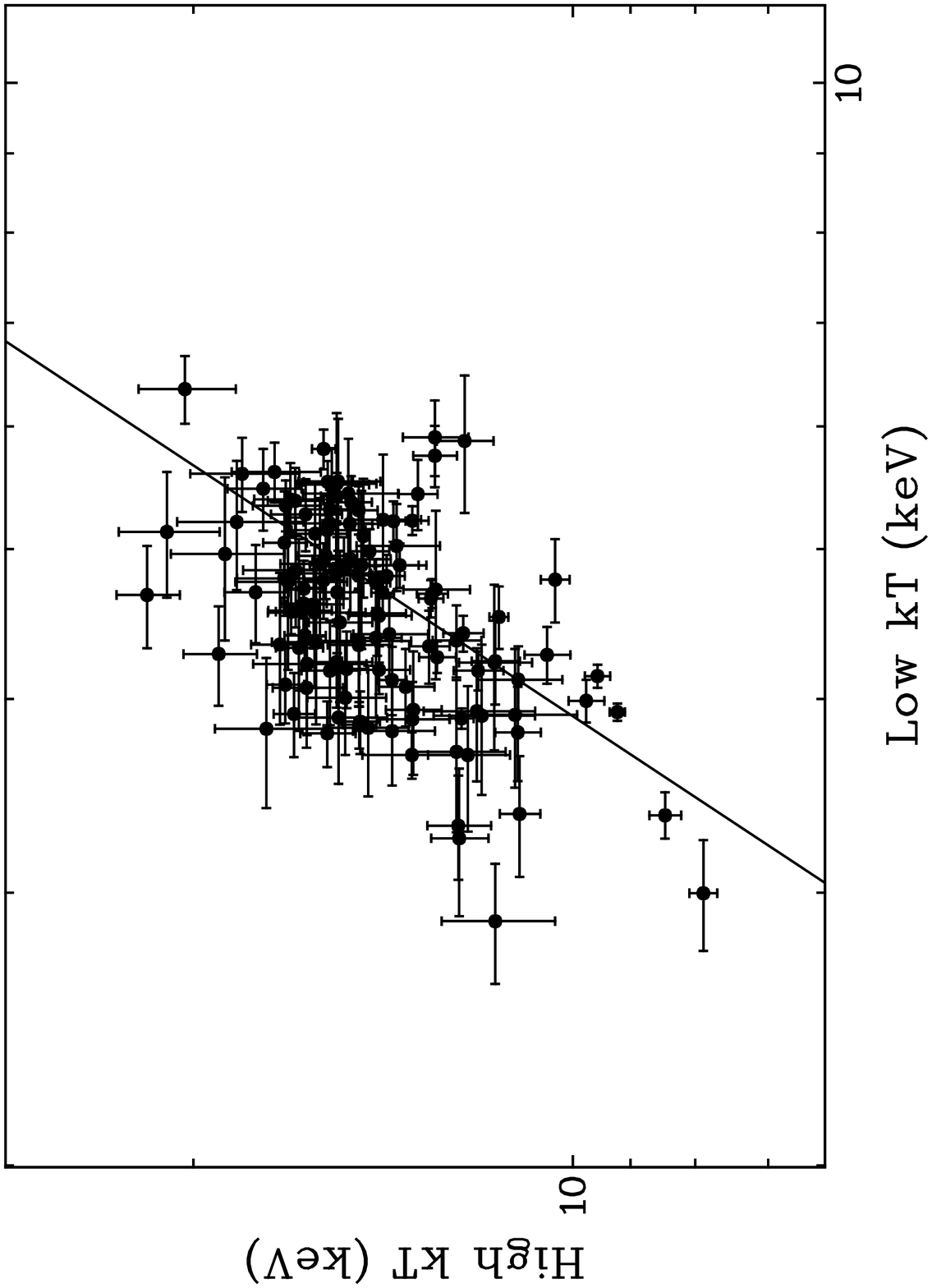}
\includegraphics[angle=-90,width=0.33\textwidth]{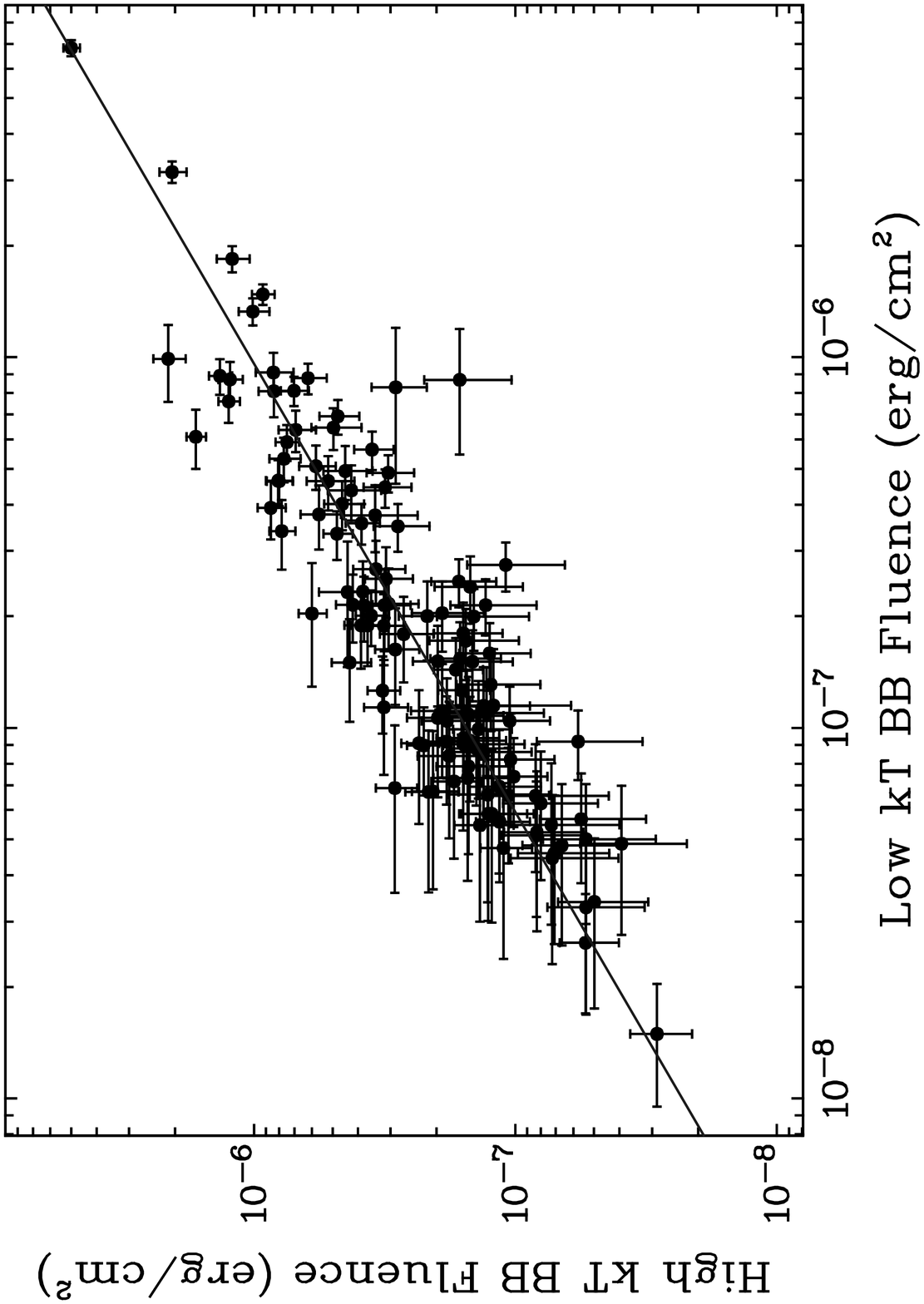}
\includegraphics[angle=-90,width=0.33\textwidth]{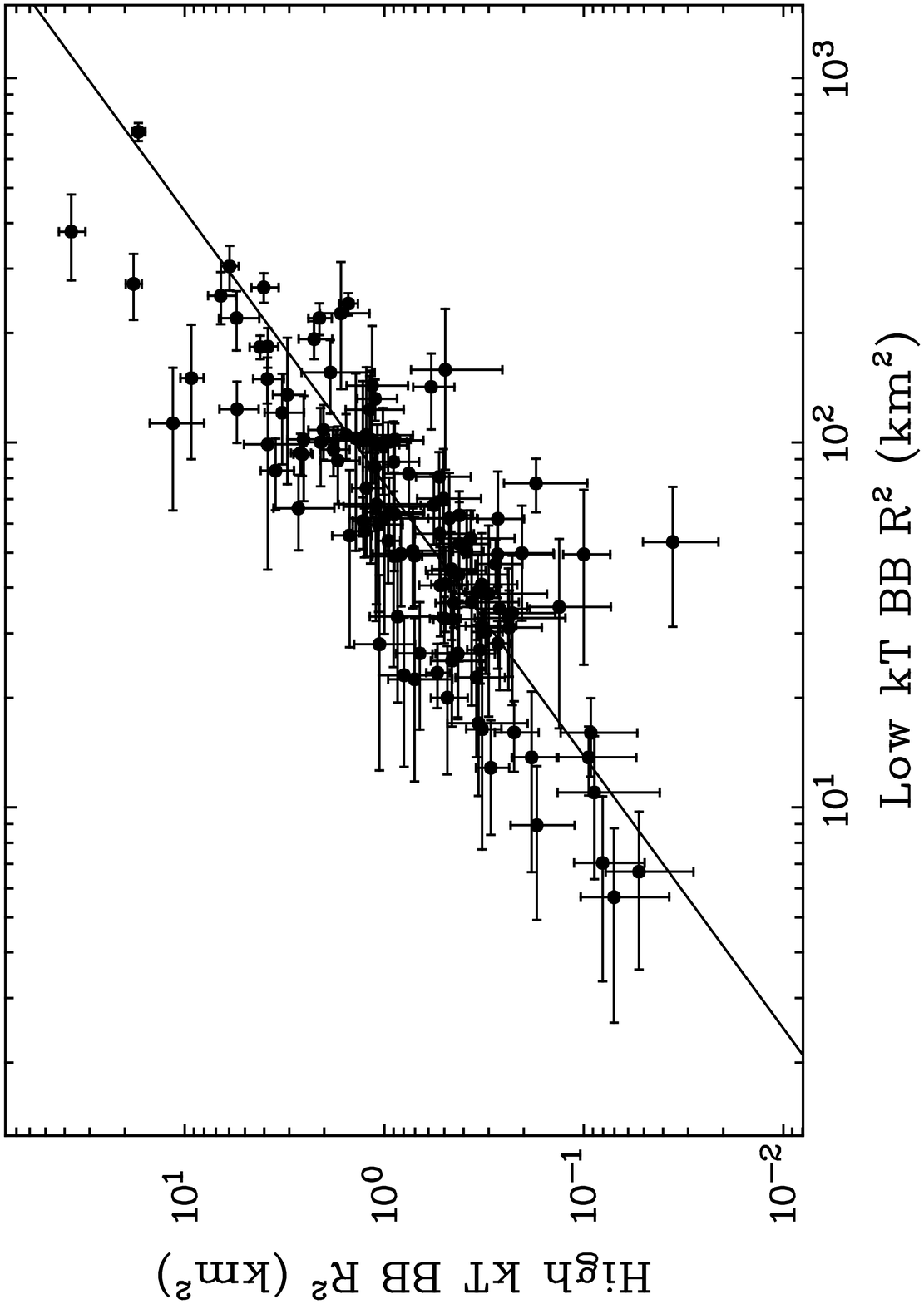}
\caption{Correlations between the cool and hot blackbody parameters: 
temperatures (left panel), fluences (middle panel), and emission areas (right panel).}
\label{fig:correl2bb}
\end{center}
\end{figure*}

The first two panels in Figure~\ref{fig:ottbcomp} show the $E_{\rm{peak}}$ distributions for the OTTB (left panel) and the Comptonized (middle panel) models. We note that these are very similar with mean $E_{\rm{peak}}$ values of 39~keV and widths of 13~keV in both cases. The OTTB function has been applied successfully to SGR bursts detected with instruments 
that have a similar or smaller spectral coverage to GBM \citep[e.g.][]{gogus1999,gogus2000,gogus2001}. 
Several studies, however, have shown that if the spectral range is extended down to several keV, 
OTTB overestimates the flux at lower energies \citep[e.g.,][]{fenimore1994,feroci2004}. 
Figure~\ref{fig:spectra} indicates a small trend for overestimation below 10~keV, 
and a similar hint is seen in several other bright bursts in our sample. 
Figure~\ref{fig:spectra} also shows that a Comptonized function with an index of $\sim-0.7$ corrects for this possible trend. 
For the brightest bursts pulse pile-up could affect the burst hardness, which would be reflected in the index value. 
We have performed simulations to study this effect, and for the count rates observed in our brightest bursts, 
the Comptonized index becomes flatter by at most $0.1$, while the $E_{\rm{peak}}$ value 
does not change significantly. 
Given that the GBM calibration \citep{bissaldi2009} was done for energies starting 
at $\sim14$~keV and then extrapolated to lower energies, at this point we cannot determine the significance of a Comptonized index deviation of 0.3 from $-1$. We thus compare in Section~\ref{sec:disc} the spectral parameters derived with both models to other SGR sources.

\subsection{Two blackbody Model}

As an alternative to OTTB or Comptonized, several studies have shown that SGR bursts can be fit well 
with two blackbody functions \citep[e.g.][]{feroci2004,olive2004,nakagawa2007,israel2008,esposito2008,lin2011}. 
We have fit 263 of our bursts in our sample (excluding the 23 saturated events) with BB+BB. Several of these, however, were faint and their parameters could not be well constrained. 
Therefore, in the following we have limited our sample to those bursts 
for which the temperatures and fluences of both black bodies have values that are at least 
two times larger than their uncertainties, resulting in a final sample size of 123 bursts. 
The left panel of Figure~\ref{fig:correl2bb} shows the correlation of temperatures for the low and high temperature BB; the average values (and widths) of the log-normal distributions for the cool and hot BB are 4.6 (0.7) and 14.8 (2.1)~keV, respectively. 

Figure~\ref{fig:correl2bb} (middle and right panels) also shows the correlations between the 
fluences and emission areas of both BBs. We estimated the significance of these correlations with a Spearman Rank Order Correlation test. In the following we designate a correlation to be significant if the chance probability, $P<5.7\times10^{-7}$, 
corresponding to $5\sigma$, when the correlation coefficient follows a normal distribution; 
a correlation is marginally significant if $5.7\times10^{-7}<P<2.7\times10^{-3}$ ($3\sigma$). 
While the temperature correlation is almost significant at the $5\sigma$ level, 
with $P=7\times10^{-7}$, 
the fluences and emission areas are very strongly correlated. 
When fitted by a power law, the slopes of the correlations are $1.86\pm0.09$, 
$0.83\pm0.02$ and $1.34\pm0.04$, for all three panels, respectively. We further discuss these correlations in Section~\ref{sec:disc}.

\begin{figure}
\begin{center}
\includegraphics[angle=-90,width=0.99\columnwidth]{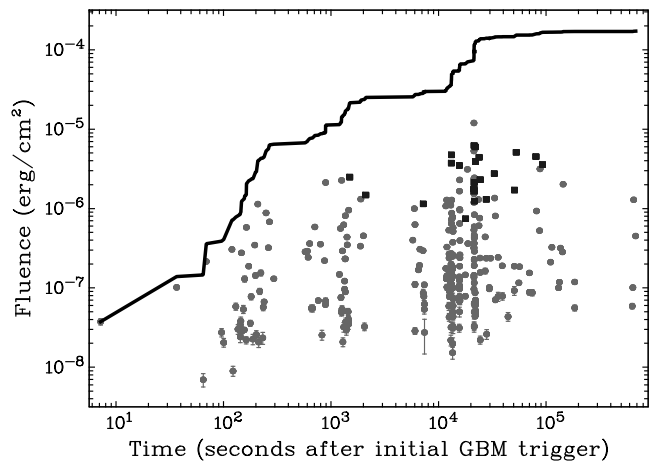}
\caption{Cumulative distribution (solid line) of the energy fluence for all 286 bursts in our sample. 
The grey data points are individual burst fluences and the black squares indicate bursts that are partially affected by saturation.}
\label{fig:cumfluence}
\end{center}
\end{figure}

\section{Discussion}\label{sec:disc}

We discuss here the energetics of the \sgr bursts using our spectral fitting results. 
We describe the correlations between the spectral and temporal parameters, 
and compare our results to the ones of other sources in the literature. 
Finally, we expand on the interpretation of our BB+BB model results.

\subsection{Burst Energetics}

We have determined the energy fluence for all the bursts in our sample in the $8-200$~keV energy range. 
The fluence values obtained with our three preferred models (OTTB, Comptonized and BB+BB) are very similar. In the following we present the fluences for the Comptonized fits only, because the model has one less parameter than BB+BB, and there are several events with power-law indices deviating from $-1$ (corresponding to OTTB). Figure~\ref{fig:cumfluence} shows the cumulative energy fluence during the source burst active period as well as the individual burst fluences. To obtain the best estimate of the total fluence released during that period in bursts, we have used data from all 286 bursts, including saturated events (black squares) whose values are merely lower limits.  
The fluence range spanned in our sample is rather large, $\sim10^{-8}$ to $\sim10^{-5}$~erg/cm$^2$, which is comparable to the range of SGR\,J0501+4516 bursts observed with GBM \citep{lin2011} and larger than previous studies of other magnetars \citep{gogus2001,gavriil2004}. 
This fluence range corresponds to an energy range in bursts of $\sim3\times10^{37}\,d_5^{\,2}$ 
to $\sim3\times10^{40}\,d_5^{\,2}$~erg, which is comparable to the values found for other SGRs, but orders of magnitude higher than AXP\,1E\,2259+586 \citep{gavriil2004}. 

The total energy fluence in bursts recorded with GBM alone is $1.7\times10^{-4}$~erg/cm$^2$, 
which is in fact a lower limit given that there were saturated bursts, as well as bursts we excluded from our analysis since no TTE data were available, or bursts that GBM did not detect because of the SAA or Earth occultation. 
This results in a lower limit on the energy emitted in bursts of $5\times10^{41}\,d_5^{\,2}$~erg, which is four orders of magnitude larger than the total energy in bursts from AXP\,1E\,2259+586 in June 2002 \citep{gavriil2004}. 
The \sgr bursts detected with {\it INTEGRAL}/SPI-ACS on 22 January 2009 
had an estimated cumulative fluence of $5.2\times10^{-4}$~erg/cm$^2$ 
\citep[25~keV$-$2~MeV][]{mereghetti2009b}, i.e., a cumulative energy of $1.6\times10^{42}\,d_5^{\,2}$~erg. 
The total energy released in bursts detected with the {\it Swift} X-ray Telescope (XRT) 
was much lower, namely $1.6\times10^{40}\,d_5^{\,2}$~erg \citep[$1-10$~keV;][]{scholz2011}. 
\citet{scholz2011} argue that the energy released in bursts is smaller than the energy released in the persistent emission 
between January and September 2009 over that same energy range, i.e., $1.5\times10^{42}\,d_5^{\,2}$~erg. They further claim that this behavior is similar to the one 
observed in AXP\,1E2259+586 and not in agreement with an SGR persistent emission. 
Although the persistent emission energy release cannot be directly measured by GBM or SPI-ACS, \citet{bernardini2011} have shown that 
for that same time period, it can be at most a factor of five higher in the $13-200$~keV than in the 
$1-10$~keV range. 
The energy released in bursts during a week is thus comparable to the energy released in the persistent emission during an eight month period \citep{scholz2011}, and we conclude that \sgr 
exhibits similar energetics behavior as other SGR sources \citep{woods2004}.

\begin{figure}
\begin{center}
\includegraphics[angle=-90,width=0.99\columnwidth]{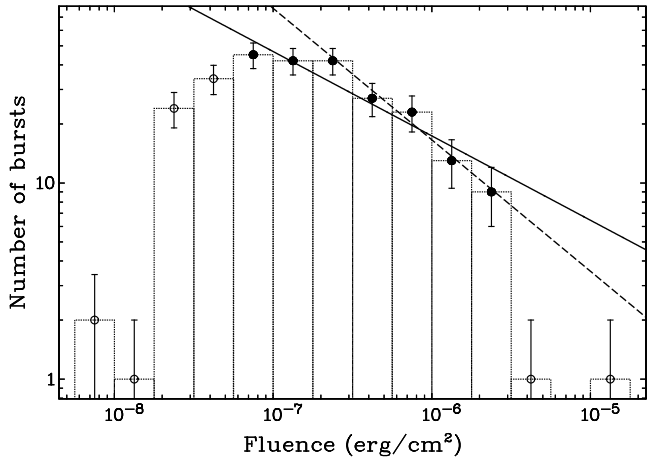}
\caption{Differential distribution of the energy fluence of \sgr bursts on a logarithmic scale.  
The solid line shows the best power-law fit to the solid circles, ranging from $10^{-7.25}$ to $10^{-5.50}$~erg/cm$^2$, with an index of $-0.4\pm0.1$. 
The dashed line shows the fit of fluences ranging from $10^{-6.75}$ to $10^{-5.50}$~erg/cm$^2$, with a power-law index of $-0.7\pm0.2$. 
The histogram error bars indicate $\sqrt{N}$ with $N$ the number of bursts per fluence histogram bin. 
Note here that the lower fluence turn-over may be reflecting either instrumental sensitivity or an intrinsic source property, or a combination of both. Therefore we give a range for the power-law slopes.}
\label{fig:energetics}
\end{center}
\end{figure}

We plot in Figure~\ref{fig:energetics} 
the differential fluence distribution for the 263 unsaturated bursts. 
The distribution is fitted, using unbinned maximum likelihood, between $10^{-7.25}$ and $10^{-5.50}$ with a power-law of index $-0.4\pm0.1$, and between $10^{-6.75}$ and $10^{-5.50}$ with an index $-0.7\pm0.2$. 
This corresponds to $dN/dF\propto F^{-1.4}$ and $dN/dF\propto F^{-1.7}$, respectively. 
The turn-over in the distribution at the low-fluence end is most likely caused by instrumental sensitivity drop-off; the onset of the turn-over is not well-determined and it depends on the criteria of the untriggered burst search algorithm. The high-fluence end of the distribution reflects that our sample is not complete with high-fluence bursts. 
When only taking the fluences above $10^{-6.75}$~erg/cm$^2$, the power-law index is comparable to the values of $\sim-0.7$ found for some other magnetar sources \citep{woodsthompson2006,gavriil2004}. 
The shallower value we obtain by including fluences down to $10^{-7.25}$~erg/cm$^2$ is closer to the values for SGR\,1806-20 \citep{gogus2000} and SGR\,J0501+4516 bursts \citep{lin2011}.

\subsection{Spectral and Temporal Correlations}

\begin{table}
\begin{center}
\caption{Spearman Rank Order Correlation coefficients and probabilities for various spectral and temporal parameters}
\label{tab:correlations}
\renewcommand{\arraystretch}{1.25}
\begin{tabular}{|l|l|l|}
\tableline
Parameters & Coefficient & Probability \\
\tableline
$E_{\rm{peak}}$ $-$ Fluence \footnotemark[a] & $-0.444$ & $3.7\times10^{-14}$ \\
T$_{90}$ $-$ Fluence \footnotemark[b] & $0.229$ & $1.8\times10^{-4}$ \\
T$_{50}$ $-$ Fluence & $0.121$ & $4.9\times10^{-2}$ \\
$\tau_{90}$ $-$ Fluence \footnotemark[a] & $0.470$ & $6.0\times10^{-16}$ \\
$\tau_{50}$ $-$ Fluence \footnotemark[b] & $0.236$ & $1.1\times10^{-4}$ \\
$\delta_{90}$ $-$ Fluence \footnotemark[a] & $0.391$ & $4.6\times10^{-11}$ \\
$\delta_{50}$ $-$ Fluence \footnotemark[b] & $0.193$ & $1.6\times10^{-3}$ \\
T$_{90}$ $-$ $E_{\rm{peak}}$ & $-0.071$ & $0.25$ \\
T$_{50}$ $-$ $E_{\rm{peak}}$ & $0.029$ & $0.64$ \\
$\tau_{90}$ $-$ $E_{\rm{peak}}$ \footnotemark[b] & $-0.209$ & $6.5\times10^{-4}$ \\
$\tau_{50}$ $-$ $E_{\rm{peak}}$ & $-0.086$ & $0.16$ \\
$\delta_{90}$ $-$ $E_{\rm{peak}}$ \footnotemark[b] & $-0.252$ & $3.4\times10^{-5}$ \\
$\delta_{50}$ $-$ $E_{\rm{peak}}$ \footnotemark[b] & $-0.260$ & $1.9\times10^{-5}$ \\
\tableline
\end{tabular}
\footnotetext[a]{Significant correlation}
\footnotetext[b]{Marginally significant correlation}
\end{center}
\end{table}

\begin{figure*}
\begin{center}
\includegraphics[angle=-90,width=0.33\textwidth]{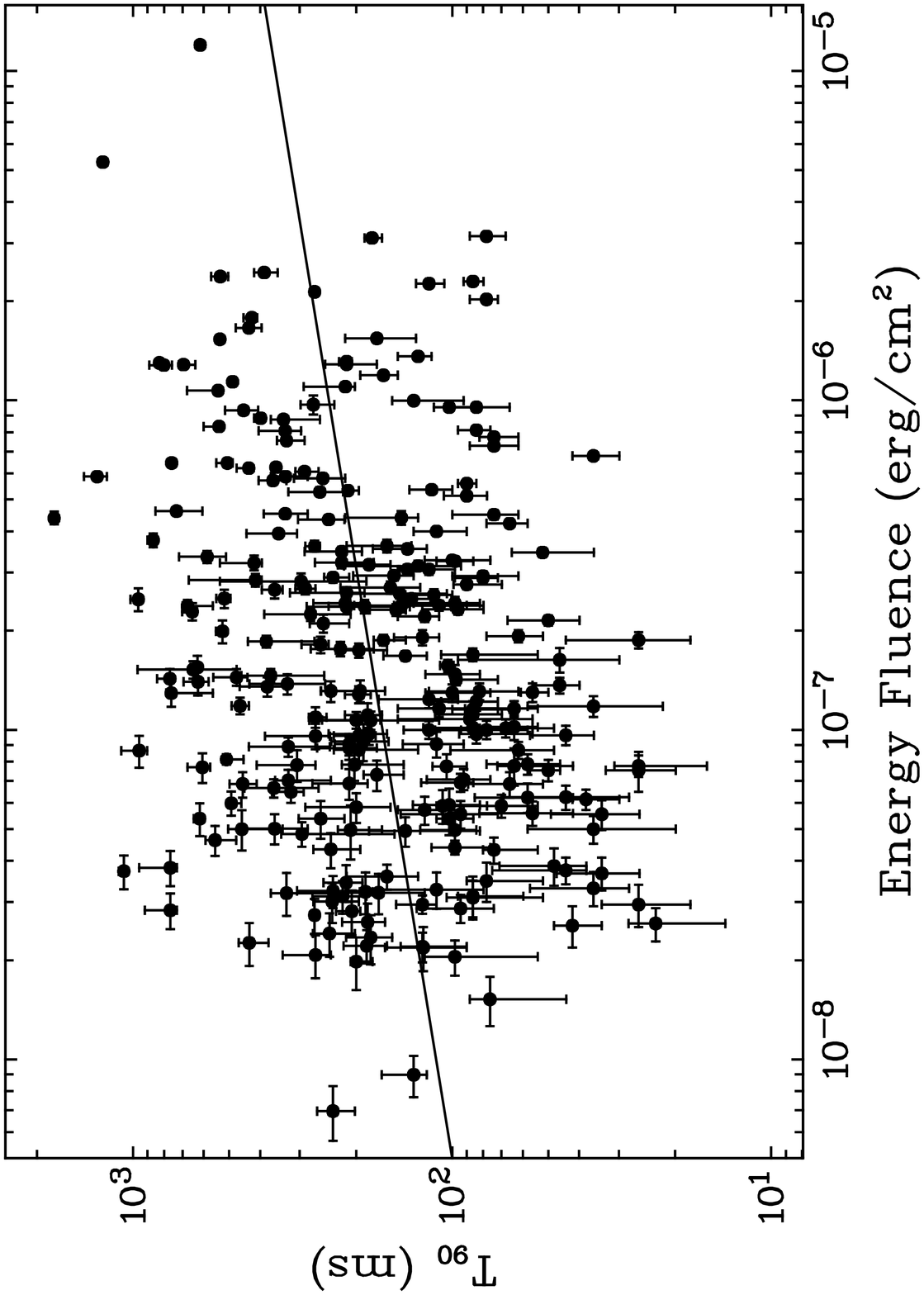}
\includegraphics[angle=-90,width=0.33\textwidth]{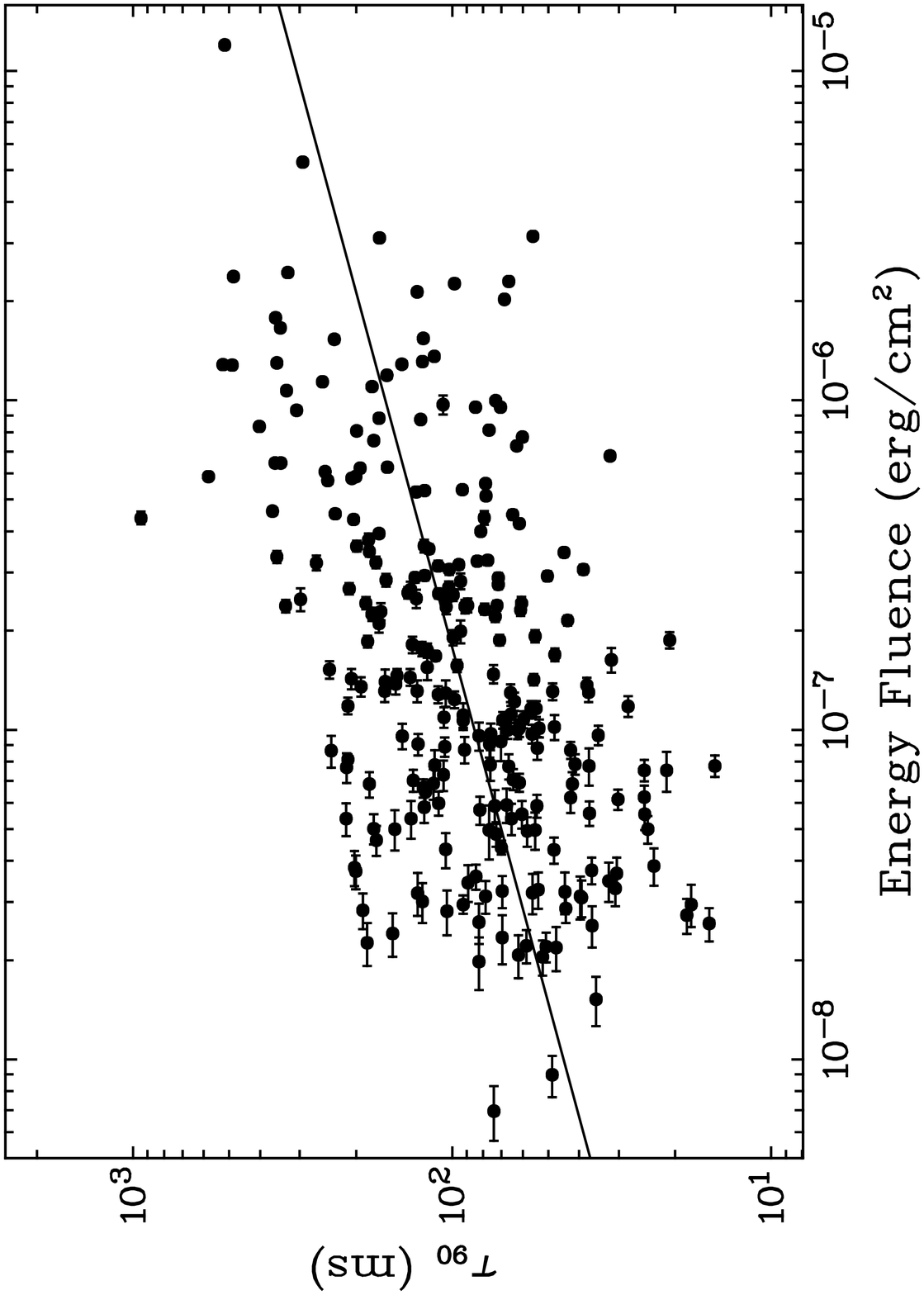}
\includegraphics[angle=-90,width=0.33\textwidth]{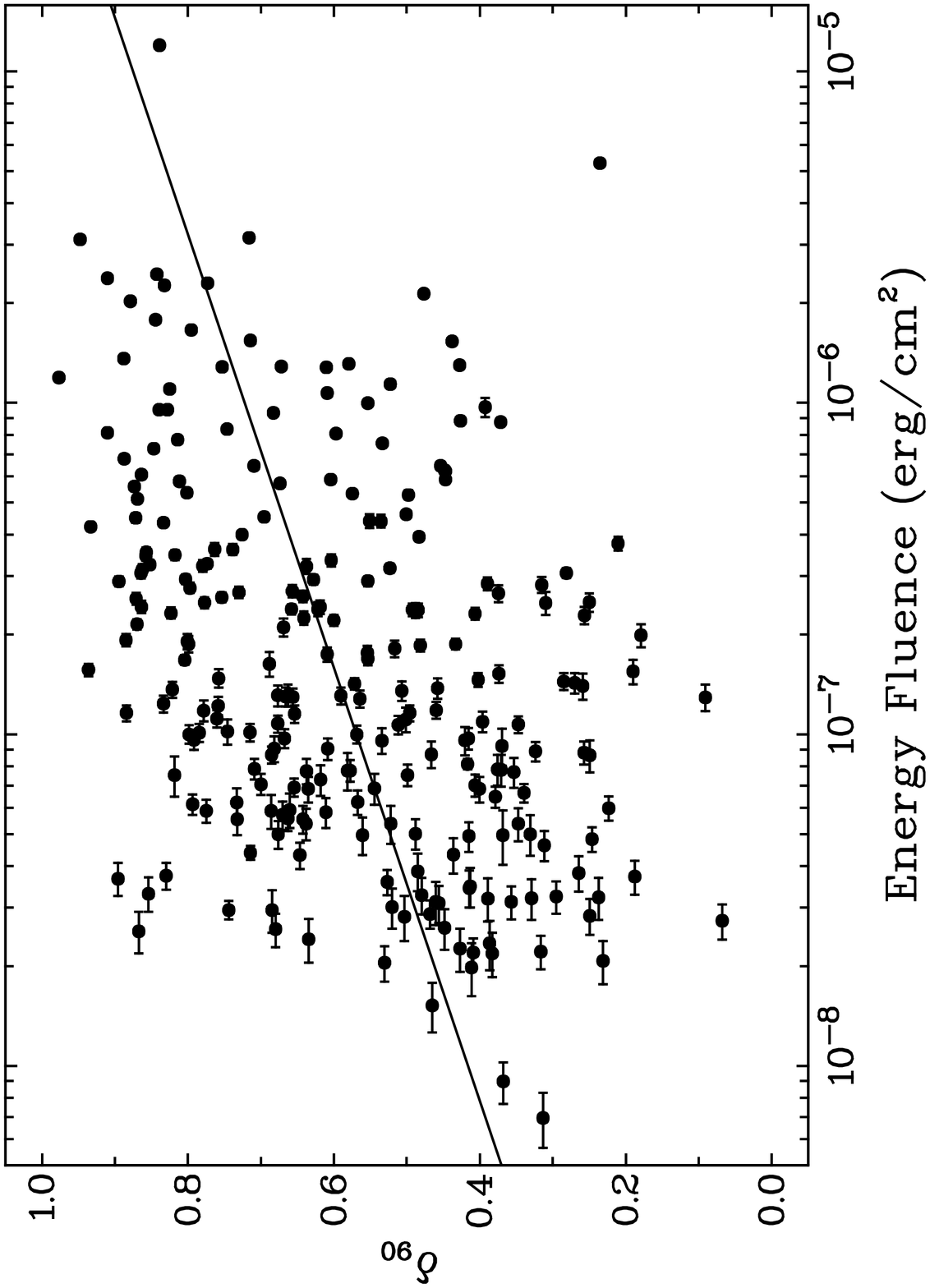}
\caption{Correlation plots of T$_{90}$, $\tau_{90}$ and $\delta_{90}$ versus energy fluence. 
The lines indicate power-law fits to the data.}
\label{fig:tempfluence}
\end{center}
\end{figure*}

Several authors have investigated correlations between the various spectral and temporal parameters 
of SGR bursts \citep[e.g.][]{gogus2001,gavriil2004}. Here we again present analysis using only the results of our Comptonized model fits. 
Using the Spearman Rank Order Correlation test we have searched for possible correlations between 
$E_{\rm{peak}}$, index, fluence, 
and the temporal parameters T$_{90,50}$, $\tau_{90,50}$ and $\delta_{90,50}$. 
There are no significant correlations between the index and any other parameter; 
for all the other correlations the Spearman Rank Correlation coefficients and chance probabilities 
are listed in Table~\ref{tab:correlations}. 
From Table~\ref{tab:correlations} it follows that the T$_{90}$ duration 
is marginally correlated with fluence. 
The emission time $\tau_{90}$ is strongly correlated with fluence and the same is true for $\delta_{90}$, 
while the correlations for $\tau_{50}$ and $\delta_{50}$ are only marginal. 
Figure~\ref{fig:tempfluence} shows the correlations of T$_{90}$, $\tau_{90}$ and $\delta_{90}$ versus fluence; 
it is apparent that there is a stronger correlation for $\tau_{90}$ than for T$_{90}$ or $\delta_{90}$. 
When fitting a power law to these correlations, we obtain indices of 
$0.17\pm0.04$ for T$_{90}$, $0.28\pm0.03$ for $\tau_{90}$, and $0.11\pm0.02$ for $\delta_{90}$.

\begin{figure}
\begin{center}
\includegraphics[angle=-90,width=0.99\columnwidth]{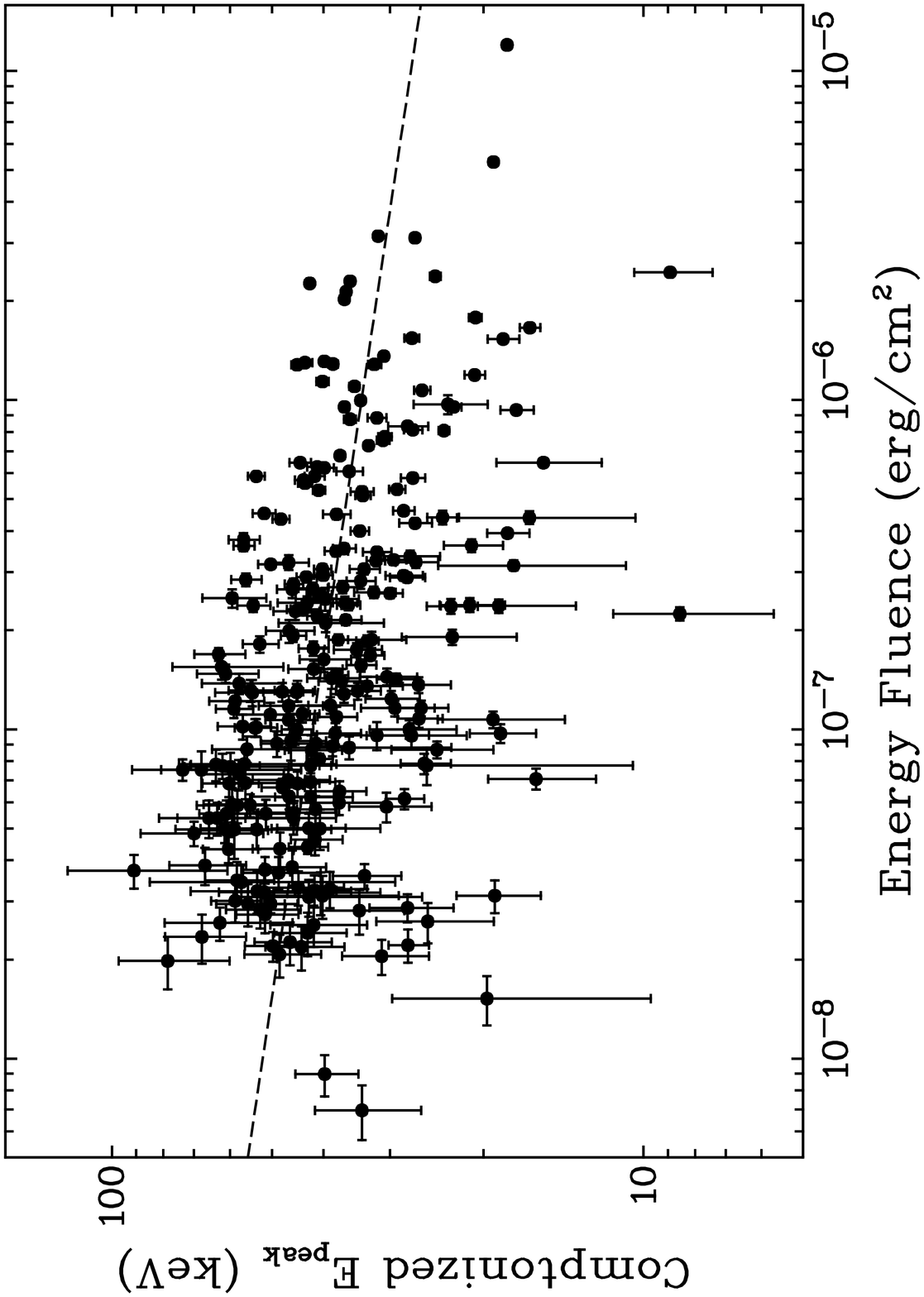}
\includegraphics[angle=-90,width=0.99\columnwidth]{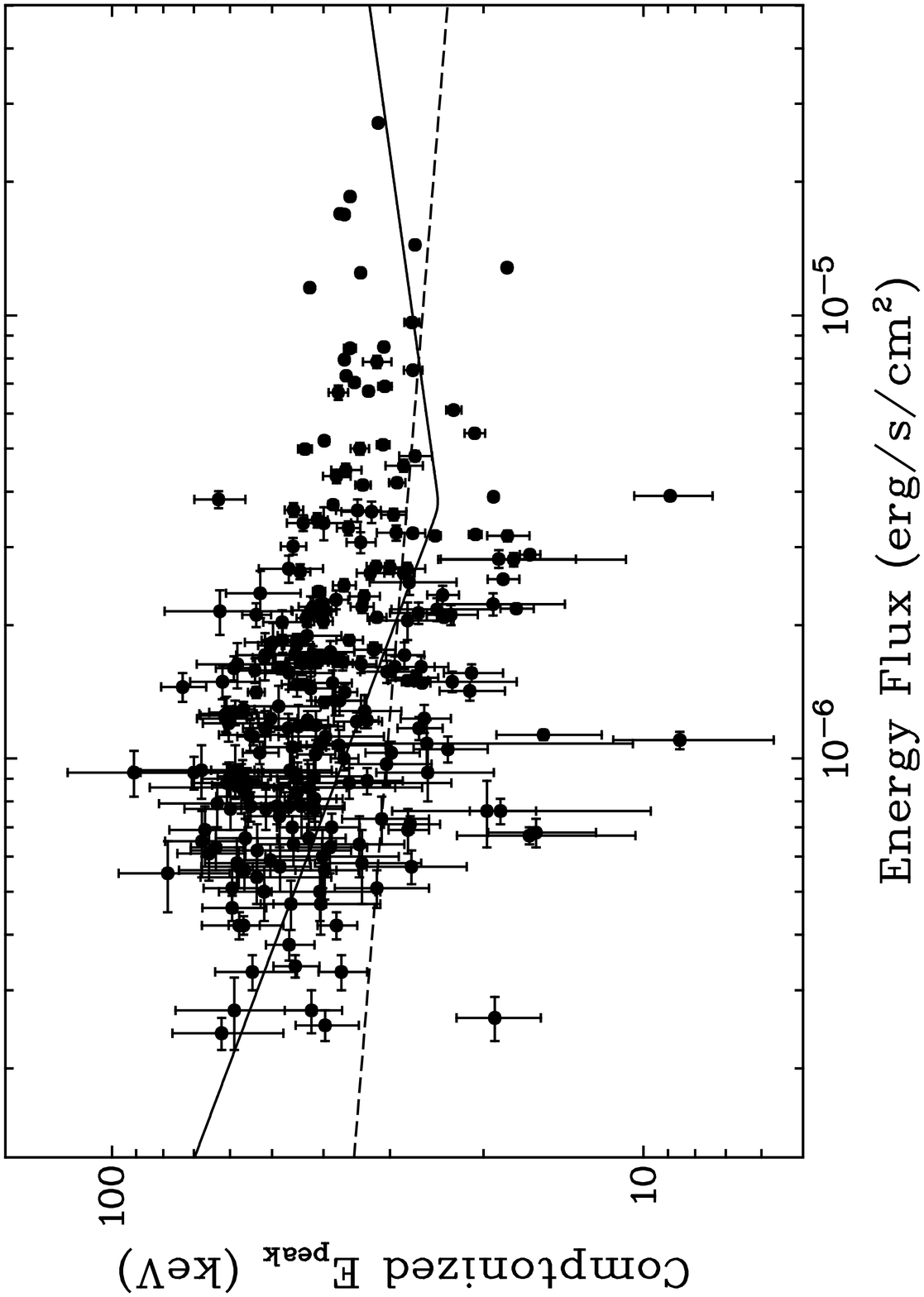}
\caption{Correlation plot between Comptonized $E_{\rm{peak}}$ and energy fluence (top panel) 
and average energy flux (bottom panel). 
The dashed lines indicate a power-law fit, while the solid line in the bottom panel indicates 
the result of a broken power-law fit.}
\label{fig:compfluence}
\end{center}
\end{figure}

There is no significant correlation between $E_{\rm{peak}}$ and the duration or emission time, 
and there is a marginal anti-correlation between $E_{\rm{peak}}$ and the duty cycle. 
$E_{\rm{peak}}$, however, is strongly anti-correlated with fluence, 
which is also clear from the top panel of Figure~\ref{fig:compfluence}. 
The bottom panel of that Figure shows a significant anti-correlation between $E_{\rm{peak}}$ 
and the average burst flux, with a Spearman Rank Correlation coefficient of $-0.344$ 
and a chance probability of $9.4\times10^{-9}$. 
Since the Comptonized index does not depend significantly on the fluence of the bursts, 
$E_{\rm{peak}}$ is a good indicator of the hardness of bursts. 
Therefore, we can conclude that there is an anti-correlation between hardness and fluence, 
and between hardness and average flux, for the \sgr bursts. 
When we fit $E_{\rm{peak}}$ versus fluence with a power law, we obtain an index of $-0.093$, 
while an $E_{\rm{peak}}$ versus average flux fit gives an index of $-0.042$ (dashed lines in both panels of Figure~\ref{fig:compfluence}). 

The bottom panel of Figure~\ref{fig:compfluence} indicates that there may be a more complicated trend 
for $E_{\rm{peak}}$ versus average flux than a single power law.
Similar to what has been found for the time-resolved spectroscopy 
of SGR\,J0501+4516 bursts detected with GBM \citep{lin2011}, we find an anti-correlation at low flux values, but a positive correlation at high flux values. 
We fit this trend with a broken power law with indices of $-0.22$ and $0.07$, with the turning point occurring at 
an $E_{\rm{peak}}$ of 30~keV and at an average flux of $4\times10^{-6}$~erg/s/cm$^2$. An F-test gives an improvement of 4$\sigma$ (chance probability of $8\times10^{-5}$) for a broken power law compared to a single one. 
We note that a Spearman Rank Order Correlation test for the fluxes below $4\times10^{-6}$~erg/s/cm$^2$ gives a coefficient of $-0.256$ and a chance probability of $7.2\times10^{-5}$, i.e. $\sim$4$\sigma$ significance, while for fluxes above $4\times10^{-6}$~erg/s/cm$^2$ there is not a significant positive correlation with a coefficient of $0.085$ and probability of $0.64$.

The minimum average flux defined by the broken power-law fit is a factor of two lower than the corresponding flux found for SGR\,J0501+4516 by \citet{lin2011}. 
Given that SGR\,J0501+4516 has an estimated distance of $\sim2$~kpc, 
the minimum of an $E_{\rm{peak}}$-luminosity correlation is three times larger 
for \sgr than for SGR\,J0501+4516. 
However, given the uncertainties in the source distances 
and in the determination of the minima in the correlations, these differences are not significant. 
Furthermore, these correlations were obtained from time-resolved spectroscopy 
in the case of SGR\,J0501+4516, while we present here only time-integrated results for \sgrnos; 
time-resolved analysis of the brightest \sgr bursts is part of a future study. 
Note that in contrast with \citet{lin2011}, in the $E_{\rm{peak}}$-fluence correlation a broken power law 
does not give an improvement over a single power law. 

The correlation or anti-correlation between hardness and brightness of magnetar bursts 
has been discussed by several authors \citep[e.g.][]{gogus2001,gavriil2004,gotz2004}.
 The spectral range and temporal capabilities of GBM allow us 
to characterize the burst spectral hardness with $E_{\rm{peak}}$, as was also done in \citet{lin2011}. 
This hardness indicator is much better defined than hardness ratios or indices of single power-law spectral fits. 
Further, our bursts span a larger fluence and flux range than any other studies with comparable sample size. 
Using large samples of bursts from SGRs~1806-20 and 1900+14 observed with {\it RXTE}, 
\citet{gogus2001} have shown that there is an anti-correlation between hardness and brightness based on hardness ratios 
between $10-60$ and $2-10$~keV. The same anti-correlation was found by \citet{gotz2004} who performed a similar analysis on {\it INTEGRAL}/IBIS data for SGR~1806-20 using hardness ratios between $40-100$ and $15-40$~keV.
Contrary to the above results, \citet{gavriil2004} showed a positive correlation for bursts of AXP\,1E2259+586 detected with {\it RXTE}, also using hardness ratios, and concluded that this behavior distinguishes AXPs from SGRs. 

For \sgrnos, \citet{savchenko2010} found a positive correlation between hardness and brightness 
based on {\it INTEGRAL}/SPI-ACS hardness ratios between $60-200$ and $20-60$~keV. 
However, \citet{scholz2011} do not find a hardness-fluence correlation 
in the {\it Swift}/XRT bursts on 22 January 2009. 
They then fit a single power law to the burst spectra, 
and find a significant anti-correlation between the index and flux, indicating a positive correlation between hardness and brightness. 
Based on this correlation, \citet{scholz2011} have argued for an AXP nature of \sgrnos. 
We discuss below why the results from \citet{savchenko2010} and \citet{scholz2011}
are in apparent disagreement with our $E_{\rm{peak}}$-fluence and $E_{\rm{peak}}$-flux anti-correlations. 
The energy ranges used by \citet{savchenko2010} and \citet{scholz2011} bracket 
the energy ranges used by \citet{gogus2001} and \citet{gavriil2004}. 
It is not trivial to compare hardness ratios over different energy ranges, 
given that the SGR burst peak emission is in the low energy band for \citet{savchenko2010} 
and in the high energy band for \citet{gogus2001} and \citet{gavriil2004}. 
More importantly, we have shown here that a single power law does not provide a good description for the \sgr burst spectra, 
which makes the method of \citet{scholz2011} less effective than using $E_{\rm{peak}}$ as a hardness indicator. Taking all these arguments into account, we conclude that the GBM hardness-fluence anti-correlation of \sgr strongly indicates that the source is very similar to other SGR sources.

\subsection{Comparison with Other SGR Sources}

\subsubsection{OTTB and Comptonized Model}

Since OTTB or Comptonized models have very often been used as the best models for SGR bursts, 
we now compare our spectral fit parameters with those from studies of other SGR sources. 
Our $E_{\rm{peak}}$ distributions (Figure~\ref{fig:ottbcomp}) are very typical for SGR bursts, 
but there seems to be a difference across SGR sources 
regarding the power-law indices of the Comptonized function. 
The average power-law index for \sgr bursts ($-0.95$) is significantly different than the one found by \citet{lin2011}
for a sample of 29 bursts detected by GBM from SGR\,J0501+4516, namely $-0.32$.  
Although the width of the index distribution is fairly large for SGR\,J0501+4516, 
$0.9$ compared to $0.41$ for \sgrnos, \citet{lin2011} show that 
their brightest bursts have 
well-constrained indices close to $0$, significantly deviating from $-1$. 
\citet{feroci2004} have fit 10 short bursts detected by {\it BeppoSAX} from SGR\,1900+14 
with a Comptonized function, and they find a mean spectral index of $-0.4$ with a dispersion of $0.2$. 
Their result is comparable to the SGR\,J0501+4516 bursts and not consistent with our values for \sgrnos. 
The spectral index differences between various SGR sources could be due to 
differences in magnetic field strength, geometry, plasma temperature or opacity. 
\citet{lin2011} discuss in detail some of the effects the extreme magnetic fields of magnetars have 
on the Comptonized power-law index.  
More detailed studies, both theoretically and observationally, 
are needed to reach conclusive results.

\subsubsection{Two blackbody Model}

\begin{figure}
\begin{center}
\includegraphics[angle=-90,width=0.99\columnwidth]{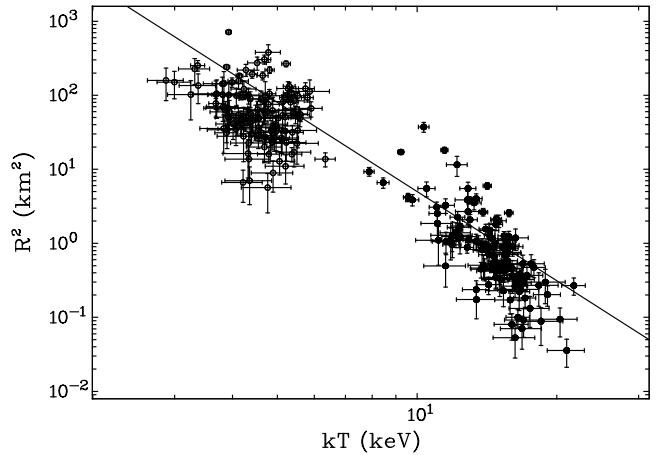}
\caption{Emission area as a function of blackbody temperature for those bursts in our sample 
that have well constrained parameters in a BB+BB fit. 
The solid line indicates $R^2\propto(kT)^{-4}$ (see text for a discussion).}
\label{fig:areatemp2bb}
\end{center}
\end{figure}

The temperatures, fluences and areas we find for the BB+BB model are similar to those of other SGR sources \citep[e.g.,][]{feroci2004,olive2004,nakagawa2007,israel2008,esposito2008,lin2011}. 
We have found a positive correlation between the emission areas of the two black bodies (Figure~\ref{fig:correl2bb}), 
which is more significant than the one found for bursts from SGRs\,1806-20 and 1900+14 \citep{nakagawa2007}. 
A strong correlation between the two BB luminosities has been found for SGR\,1900+14 bursts with {\it Swift} 
\citep{israel2008} and for SGR\,J0501+4516 bursts with GBM \citep{lin2011}. 
The latter two studies, however, have used time-resolved spectral analyses, 
while we show here a strong fluence correlation between the two BBs for time-integrated spectra. 

Figure~\ref{fig:areatemp2bb} displays the emission area as a function of temperature for both BBs. 
The temperature of the cool BB does not show any correlation with the fluence or the emission area of that BB. 
The hot BB does have a marginal anti-correlation between temperature and fluence, 
with a chance probability of $1\times10^{-3}$ in a Spearman Rank Correlation test, 
and a very strong anti-correlation between the temperature and the emission area 
(Figure~\ref{fig:areatemp2bb}) with a chance probability of $3\times10^{-24}$. 
For illustrative purposes we have drawn a solid line in Figure~\ref{fig:areatemp2bb} indicating $R^2\propto(kT)^{-4}$, 
which corresponds to a constant burst luminosity or fluence. 
The line seems to represent best the low-temperature part of the hot BB data, while
there is a clear steepening at the higher temperatures. 
Indeed, a power-law fit of the hot BB data gives a slope of $\sim-7$, 
similar to what has been shown for SGR\,J0501+4516 by \citet{lin2011}. Similar studies have been performed for SGRs\,1900+14 \citep{israel2008} and 1627-41 \citep{esposito2008} using time-resolved spectroscopy; our results confirm the trends described in these studies. A detailed time-resolved spectroscopy of \sgr bursts is underway. 

Finally, we note that the smallest emission areas of the hot BB (few hundredths of km$^2$) 
are comparable to the emission area of the BB component 
found during enhanced persistent emission from \sgr in the GBM data 
after the bursting onset on 22 January 2009 \citep{kaneko2010}, 
possibly indicating a common origin of the two phenomena. 
For the outburst data studied here, the compact hot blackbody component
can be as small as $R \sim 0.2-0.3\;$km in size, corresponding to a
diameter of the order of the thickness of the outer crust, yet the bulk
of the $R$-values for the hot BB component correspond to a diameter of
$\sim 1\;$km -- comparable to the total thickness of the solid crust.
For energy injection at or just below the surface, if these scales
signify the size of the region in the crust that is fractured by the
magnetic stresses, as is likely, then injection scales considerably
smaller than the crust height have greater difficulty in disrupting it
and initiating a flare, particularly if the anchored magnetic flux tube
is non-radial at its footpoint.  To leading order, the field energy that
threads the injection region scales as $R^2$, while the length of the
required fracture and the corresponding energy required to produce it
scale as $R$.  Hence, crustal disruption by field twists or shear is
expected to be less effective for smaller $R$, a contention that appears
to be borne out in the data: the majority of bursts possess $R$ values
for the hot component on the scale of the thickness of the crust.
Moreover, flares with a larger $R$ for the hot BB component tend to have
a larger total energy output, i.e., luminosity or fluence.

\subsection{Interpretation of Two blackbody Model Results}

The BB+BB model results point towards a photospheric interpretation, and
perhaps one with dynamic evolution, from one magnetospheric locale to
another. The fitted temperatures exceed those found ($\sim0.5$~keV) in
the classic X-ray band ($0.5-10$~keV) in quiescent magnetar emission
\citep[e.g., see][]{perna2001}, indicating a magnetospheric origin for
such photospheres, as opposed to generation of emission purely on the
neutron star surface. While it is possible that the site of outburst
activation could be near the magnetic poles, the $T_{90}$ durations of
\sgr far exceed the neutron star light crossing time. Hence, even the
intense magnetic fields of magnetars cannot restrict the emission region
of powerful outbursts to small volumes. Accordingly, one expects spatial
transport of the burst luminosity during each flare, even if tied to
closed field lines. This is essentially the picture that
\citet{duncanthompson1992} originally envisaged for SGRs.  Polar origins
would permit rapid plasma expansion along open field lines to very high
altitudes, so that plasma containment is limited. Near the equator, trapping
of the gas is optimal, enabling longer durations of emission, so that
quasi-equatorial locales for energy injection might be favored.  The
trapping times depend intimately upon the complicated interplay of
polarization-dependent Compton scattering, and associated lepton
diffusion and energy exchange within the photon-electron gas.

The core property evinced in Figure~\ref{fig:areatemp2bb} is that the
burst luminosity $\propto (kT)^4\,R^2$ of the collection of bursts (and
approximately also for individual bursts) is similar for the low and
high temperature components. This is a total energy equipartition
feature: if the surface area of a putative magnetospheric emission
volume scales approximately as $R^2$, then the energy in each black-body
component is approximately equal. 
An $R^2$ scaling most likely applies to the hotter blackbody component
as it evolves in a coronal structure, with photon and particle
propagation away from an injection zone.  It could be pertinent for
burst activation locales near the magnetic poles, but also if the
injection site is somewhat remote from the surface, for example in the
magnetospheric twist scenario of \citet{thompson2005} and
\citet{beloborodov2007}, which was primarily envisaged for the lower
luminosity, quiescent magnetar emission.  However, it is possible that
other volume scalings are operating, particularly due to the
anisotropizing influence of the non-uniform magnetic field.  For
example, volumetric constriction of hot plasma near field line
footpoints might modify this correlation modestly. Determining the
effective volume scaling requires detailed time-dependent modeling of
particle and radiative transfer in a magnetospheric environment.  This
motivates future theoretical work on magnetar ``fireballs.''  Yet, for
the present, a baseline conclusion is that it is clear that the strong
correlation between the energies in the two components suggests that
they share the same physical origin if BB+BB is indeed the correct model for these bursts. 

It is natural to expect that the high temperature component would be
more closely connected to a smaller injection region, where energy is
dissipated from crustal stress fractures
\citep{duncanthompson1992,thompsonduncan1995} or some other origin.
Proximity of the injection site to the surface is favored in such a
scenario, and the resultant fireball of electrons and photons should
subsequently expand and cool at higher altitudes in the closed
magnetospheric zone, thereby generating the lower temperature component.
The injection temperature is controlled by the total energy dissipated
by hot electrons present in the inner magnetosphere. The magnetic
Thomson optical depth should be high, so that Comptonization will drive
thermalization, and the fireball evolution and duration will be
influenced by the anisotropic, magnetic Eddington luminosity.
Equilibration between photons and electrons will be non-uniform over the
coronal volume, so that there should be a modest temperature gradient
throughout; the emergent continuum will be a superposition of distorted
blackbodies. In particular, the strong polarization dependence of the
anisotropic Compton process naturally establishes two different physical
scales for the optically thick environment \citep{ulmer1994,miller1995},
with the so-called E-mode photosphere being smaller than the O-mode
photosphere. Perhaps these physically-distinct regions assume somewhat
different temperatures on average, though a more quantitative
development of this picture is beyond the scope of this
paper.  Whatever the site of the injection, one anticipates that
there will be distinctive evolutionary signatures in the dynamic
fireball as it either expands to higher altitudes, or contracts when
constricted near magnetic footpoints, resulting into a certain temporal
evolution for the temperatures. Evolutionary aspects of the spectroscopy
of \sgr bursts will be explored in a future paper.

\section{Conclusions}\label{sec:conc}
 
We have presented the results of a time-integrated spectral and temporal study of bursts from \sgr 
during an extremely active bursting period, making optimal use of the spectral and temporal capabilities of GBM. 
The durations, emission times and duty cycles are typical of SGR bursts, 
with burst rise times shorter than their decays. 
Our spectral analysis shows that the spectra of \sgr bursts are equally well described 
by OTTB, Comptonized and BB+BB. 
The $E_{\rm{peak}}$ distributions of OTTB and Comptonized have mean values of $\sim39$~keV, 
while the Comptonized power-law index average is close to $-1$, 
i.e., we recover the OTTB index in our Comptonized fits. 
Whether this power law extends down to lower energies can only be determined 
by combined fits with other instruments, for instance {\it Swift}/XRT (Lin et al., in preparation).

We have investigated the presence of correlations between the various spectral and temporal parameters, 
and find significant correlations between the emission time and the fluence, 
and between the duty cycle and the fluence, and a marginal correlation between the duration and the fluence. 
We have also shown that there is a significant anti-correlation between $E_{\rm{peak}}$ and fluence, and between 
$E_{\rm{peak}}$ and average flux. 
We find an anti-correlation at low flux levels but a weak correlation with $E_{\rm{peak}}$ at high fluxes, with a possible minimum at 4~erg/s/cm$^2$. In a time-resolved analysis of the bursts from SGR\,0501+4516, the positive correlation behavior at high fluxes is more significant. Whether the two sources have a similar behavior remains to be seen in the time-resolved analysis of the \sgr bursts.  
One of the main contributions of the GBM data is providing accurate measurements of the evolution of $E_{\rm{peak}}$ for short bursts, which is the best hardness indicator for hardness-brightness correlation studies.  

For the BB+BB fits we find average temperatures of $\sim5$ and $\sim14$ keV 
and we show that these temperatures are well correlated. 
The fluences of the two BB functions are strongly correlated, 
and the same is true for their emission areas. 
The emission area of the low-temperature BB is comparable to the neutron star surface area; the hot BB area is much smaller and strongly anti-correlated with its temperature.

\acknowledgements
{\small AJvdH would like to thank Vicky Kaspi for useful discussions. 
This publication is part of the GBM/Magnetar Key Project (NASA grant NNH07ZDA001-GLAST, PI: C. Kouveliotou). 
Y.K. and E.G acknowledge the support from the Scientific and Technological Research Council of Turkey (T\"UB\.ITAK) through grant 109T755. 
MGB acknowledges support from NASA through grant NNX10AC59A. 
SG was supported by an appointment to the NASA Postdoctoral Program at the Goddard Space Flight Center, administered by Oak Ridge Associated Universities through a contract with NASA. 
JG is supported by the ERC advanced research grant ``GRBs''. 
ALW acknowledges support from an NWO Vidi Grant.}

\end{document}